\documentclass[11pt]{article}
\usepackage[margin=1in]{geometry}
\usepackage{amsmath,amssymb,amsthm,mathtools}
\numberwithin{equation}{section}
\usepackage[numbers]{natbib}
\usepackage{etoolbox}
\makeatletter
\patchcmd{\NAT@test}{\else \NAT@nm}{\else \NAT@nmfmt{\NAT@nm}}{}%
  {\PackageError{spat}{Could not enable capitalized numeric citations}{Check the natbib version.}}
\makeatother
\usepackage[breakable]{tcolorbox}
\usepackage{enumitem}
\usepackage{graphicx}
\usepackage{xspace}
\usepackage{xcolor}
\usepackage{authblk}
\graphicspath{{figures/}}
\usepackage{xcolor}
\usepackage[colorlinks=true,citecolor=blue]{hyperref}

\renewcommand{\Re}{\mathbb{R}}
\newcommand{\E}{\mathbb{E}}
\newcommand{\ind}[1]{\mathbf{1}\{#1\}}
\newcommand{\inc}{{\mathrm{I}}}
\newcommand{\dfrd}{{\mathrm{D}}}
\newcommand{\calF}{\mathcal{F}}
\newcommand{\calR}{\mathcal{R}}
\newcommand{\calQ}{\mathcal{Q}}
\newcommand{\calC}{\mathcal{C}}

\renewcommand{\d}{\,\textnormal{d}}

\newcommand{\muhat}{\widehat{\mu}}
\newcommand{\mix}{\textnormal{mix}}
\newcommand{\eps}{\epsilon}
\newcommand{\cgf}{{\textsc{cgf}}}
\newcommand{\gro}{{\textsc{gro}}}
\newcommand{\drift}{{\textsc{drift}}}
\newcommand{\plugin}{{\textsc{plug-in}}}
\newcommand{\lambdagro}{\lambda_{\gro}}

\newcommand{\aaaibreak}[1][]{\ifarxiv\else\\\aaaiamp#1\fi}

\DeclareMathOperator*{\argmax}{\text{argmax}}
\DeclareMathOperator*{\argmin}{\text{argmin}}
\DeclareMathOperator*{\esssup}{\text{ess}\,\text{sup}}

\newtheoremstyle{spaced}%
  {\medskipamount}
  {\medskipamount}
  {\itshape}
  {}
  {\bfseries}
  {.}
  {0.5em}
  {}
\theoremstyle{spaced}
\newtheorem{theorem}{Theorem}
\newtheorem{proposition}{Proposition}

\newtheorem{lemma}{Lemma}
\newtheorem{definition}{Definition}
\newtheorem{remark}{Remark}

\newif\ifarxiv
\arxivtrue

\title{Principal-agent hypothesis testing \\ under adaptive experimentation}
\author[1]{Ben Chugg}
\author[2]{Aritra Ghosh}
\author[3]{Aaditya Ramdas}
\affil[1]{\small Carnegie Mellon University}
\affil[2]{Indian Statistical Institute}
\affil[3]{Stanford University}

\date{October 2026}

\begin{document}

	\maketitle

    \begin{abstract}
Principal-agent hypothesis testing studies statistical contracts in which a principal rewards an agent according to data the agent generates. Existing work focuses on the single-shot setting, where the agent produces a single dataset. We extend the framework to sequential settings, where evidence accumulates over time, rewards may be deferred to the end or disbursed incrementally, and the agent may stop the contract at a strategically chosen stopping time. We call a contract null-deterring if no ``null agent'' (those whose data-generating distribution lies in the null hypothesis) can profit in expectation under any stopping rule, and our main conceptual result shows that such contracts are characterized by e-processes. We then turn to the agent's best response and identify the optimal stopping rule through the Snell envelope of the reward process. We develop simpler, more interpretable approximations that often require the agent to have only partial knowledge of its distribution.
\end{abstract}

	\section{Introduction}

	Consider a drug regulator (e.g., the FDA) that controls access to a market. A pharmaceutical company can pay a cost $C>0$ to run a clinical trial of a new treatment. Based on the outcome of the trial, the regulator decides how widely the treatment may be marketed. There are competing interests at play: the regulator wants effective treatments to reach patients while keeping ineffective ones off the market, whereas the company wants its treatment authorized as widely, and as quickly, as possible.

	This scenario is an example of a \emph{principal-agent problem}: the strategic interaction between a principal, who controls access to a good, and an agent who is vying for access. Such examples are not hard to find in practice: recommendation algorithms, internal experimentation platforms at tech companies, and venture funding being a few common ones.
    Given their ubiquity, principal-agent problems have long been studied in economics (e.g., \citep{grossman1992analysis,rogerson1985first,vaubel2006principal,gerardi2012principal}).

	Recently, motivated by the FDA example above, \citet{bates2022principal} introduced a framework to analyze such dynamics from a statistical perspective, which they call principal-agent hypothesis testing.  The framework is based on \emph{statistical contracts}:  protocols which guide the interaction between the principal and agent. The agent pays a cost to enter the contract and then generates data from a distribution---a distribution that is private and unknown to the principal, but possibly known to the agent. The principal allocates resources (rewards) based on inference made from this random sample.  Importantly, the initial cost and reward mechanism was agreed to \emph{a priori} by the agent, before generating the data.

	In the example above,  the trial outcome is a random sample drawn from the agent's distribution. The principal does not know the treatment's effectiveness beforehand. The reward, perhaps a license to profit or the scope of market authorization, is a function of the observed outcomes. In general, the reward may be profit, market share, a license, a binary accept/reject decision, etc.

	Agents are assumed to be partitioned into null agents (e.g., companies with ineffective treatments) and non-null agents (those with effective treatments). Formally, there is a parameter space $\Theta$ partitioned into $\Theta_0$ (nulls) and $\Theta_1$ (non-nulls) and each agent has a private type/parameter $\theta\in\Theta$. Their data is drawn from $X\sim P_\theta$. With this setup in mind, the principal-agent hypothesis  testing framework of \citet{bates2022principal} is based on the following statistical contract:

	\begin{tcolorbox}[
		colback=blue!8!white,
		colframe=white,
		boxrule=0pt,
		sharp corners,
		left=6pt,
		right=6pt,
		top=6pt,
		bottom=6pt
		]
		\textbf{Single-shot statistical contract \citep{bates2022principal}}
		\begin{enumerate}[nosep,topsep=0.1cm]
			\item The principal publicly announces a cost $C>0$ and a menu $\calR$ of reward functions.
			\item An agent with hidden parameter $\theta$ may pay a cost $C>0$ to enter the contract. Upon entry, the agent selects a reward function $R\in\calR$.
			\item The agent then generates data $X \sim P_\theta$ and receives reward $R(X)\in[0,\infty)$.
		\end{enumerate}
	\end{tcolorbox}

	The central design problem in statistical contract theory is to control the expected rewards allocated to null agents while allowing non-null agents to earn rewards commensurate with the evidence they submit~\citep{jordan2025collectivist}. \citet{bates2022principal} show that \emph{null-deterring} contracts\footnote{They call them \emph{incentive-compatible}.}---those ensuring that null agents cannot earn a positive expected  profit---must consist of rescaled e-values, a property that follows almost directly from the definition (and motivation) of e-values. They further show that e-values characterize maximin optimal contracts, thus providing actionable advice for any principal looking to design statistical contracts.

	There is a drawback, however. The contract above assumes that a single dataset is produced by the agent. But in many applications it is natural to consider sequential experiments. Data may take time to procure, or the agent may wish to obtain partial rewards based on the results thus far.
    Clinical trials are a paradigmatic case. They run over months or years, so rather than wait for the trial to conclude, a regulator may grant accelerated or conditional approval on interim evidence and broaden the authorization as more data arrive. In fact, sequential allocation and experimentation is now pervasive in the technology industry and elsewhere (e.g., Spotify~\citep{schultzberg2023choosing}, Netflix~\citep{lindon2024sequential}, Eppo~\citep{lucht2024sequential}, Growthbook~\citep{growthbook_sequential_testing}, Abridge~\citep{oberst2024pioneering}, Adobe~\citep{adobe2025experimentation_panel}, as well as in government and public policy~\citep{henderson2022beyond,henderson2023integrating}).

	Our goal in this work is to extend the single-shot statistical contract of \citet{bates2022principal} to the sequential setting, thus enabling adaptive experimentation and interim decision-making. Our contributions are as follows:
	\begin{enumerate}
		\item We propose two versions of a sequential statistical contract: one with deferred rewards and one with incremental rewards. Extending the notion of null deterrence from the single-shot setting, we define null-deterring sequential contracts as those where a null agent's expected reward is at most the entry cost under every parameter and stopping strategy.
		\item We classify the set of null-deterring contracts under both reward schedules using e-processes. For incremental reward schedules in particular, we show how to construct such contracts using functions called \emph{adjusters}~\citep{dawid2011probability,dawid2011insuring,shafer2011test}, which also arise in preserving accumulated evidence and combining e-processes~\cite{choe2026combining}.
		\item We analyze the agent's best response under both reward schedules, deriving utility bounds and simple policies with explicit performance guarantees. We also develop approximations that require less knowledge of the agent's distribution and compare these policies experimentally.
	\end{enumerate}

	We note that \citet{bates2022principal} also gave a sequential version of the single-shot setup above, but their time-horizon must be known in advance. That is, the contract can only proceed for some fixed number of rounds determined a priori. We give more detail on the differences between their sequential contract and ours in Appendix~\ref{app:bates-comparison}, but here let us summarize three ways in which our contract differs: (i) we allow stopping rules for both the principal and the agent, (ii) we study two distinct reward schedules, and (iii) we handle unbounded horizons.
	That said, their sequential contract also has benefits over ours, such as the ability to consider per-round costs. We thus view our contribution as extending the single-shot contract in an alternative way to that originally proposed by \citet{bates2022principal}.

	Code is available at \url{https://github.com/bchugg/statcontracts}.

    \ifarxiv
    \subsection{Related work}
    \label{sec:related-work}

    Since the original article of \citet{bates2022principal}, there have been several follow-up works on principal-agent testing. \citet{chen2026strategic} study a similar setup but focus on  characterizing the principal's optimal p-value threshold under strategic participation of the agent. \citet{shi2024sharp} study the agent's optimal response under general concave utility functions, and then later examine the situation of a principal interacting with multiple heterogeneous agents and focus on the issue of adverse selection~\citep{shi2025instance}. \citet{bates2023incentive} study a Bayesian version of the principal-agent problem, and \citet{chugg2026strategic} study a multi-agent version.

    Recently, \citet{velasco2026optimizing} also studied principal-agent hypothesis testing in a sequential setting, though with different goals. In their model, the principal seeks to maximize social welfare by subsidizing a fraction of the agent's experimentation cost, and the agent is uncertain about efficacy and updates a prior private belief. Despite these differences, the two works share key ingredients: both rely on e-values~\citep{ramdas2025hypothesis}, and both characterize the agent's optimal strategy via dynamic programming. We therefore view their work as complementary to ours, and together they highlight the many possible extensions of principal-agent problems to more dynamic settings.

    \citet{nguyen2026incentive} also study sequential statistical protocols in which an agent chooses whether to submit a project. The principal observes public testing outcomes and uses future access to testing to influence these decisions. In this setup, the agent repeatedly decides whether to submit a new project, whereas ours accumulates evidence about a fixed type. \citet{velasevic2026anytime} use anytime-valid tests to design contracts that protect users of AI services from a provider covertly downgrading its model.

    Our analysis of the agent's parameter choice is closely related to recent work of \citet{clerico2026time} on sequential testing with preferences over rejection times. Their EDO criterion coincides with our CGF-optimal criterion, and our  bounds adapt their single-threshold bounds. The settings differ, however. They  choose parameters to maximize a discounted reward for rejecting the null at a single significance level. Our agent faces a contract whose reward grows with the evidence and is not always paid out all at once. The bounds must therefore hold across many evidence levels and be combined with the agent's stopping decision.

    Contract theory has a long history outside of the particular statistical framing studied here. Its primary goal is to study the interactions of agents, at least one of whom has private information. We refer to \citet{lanchimba2026contract} and \citet{laffont2002theory} for overviews.
    The adjacent field of \emph{algorithmic} (as opposed to \emph{statistical}) contract theory~\citep{dutting2024algorithmic} is also concerned with principal-agent problems, but emphasizes the computational aspect as opposed to the statistical aspect. There, the goals are to learn optimal payout vectors and to analyze the computational complexity of doing so~\citep{cohen2022learning,chen2024bounded,zhu2022sample}.

    \fi

	\subsection{Technical preliminaries}
	Let $(\Omega,\calF)$ be a measurable space.
	An e-value for a set of probability distributions $\calQ$ on $(\Omega,\calF)$ is a nonnegative random variable $E:\Omega\to [0,+\infty]$ such that $\E_Q[E]\leq 1$ for all $Q\in\calQ$. For readers unfamiliar with e-values, it is helpful to think of e-values as statistical cousins of p-values. While developed for similar purposes (in particular, to provide a measure of evidence against the null~\citep{chugg2026values}) their definition in terms of expectations instead of tail probabilities gives them useful properties that p-values do not have. We refer to \citet{ramdas2025hypothesis} and \citet{grunwald2024safe} for more.

	As our motivation in this work is to consider sequential settings, we will require the sequential version of an e-value, the e-process. To define it,
	we suppose that the space $(\Omega,\calF)$ admits a filtration $(\calF_t)_{t\geq 0}$. Then an e-process for $\calQ$ is an $(\calF_t)$-adapted sequence of nonnegative random variables $(E_t)_{t\geq 0}$ (that is, $E_t$ is $\calF_t$-measurable for all $t$) such that $E_\tau$ is an e-value for $\calQ$ at all stopping times $\tau$. Recall that a stopping time is a random variable such that $\{\tau = t\}$ is $\calF_t$-measurable for all $t\geq 0$. That is, at time $t$, one can always determine whether the stopping time is now. \ifarxiv All logarithms throughout the paper are natural logarithms.\fi

	\section{Sequential Statistical Contracts}
	\label{sec:sequential-contract}

	Our first order of business is to determine how to extend the reward menu $\calR$ in the single-shot contract to the sequential setting. Instead of a set of functions, we will have a set of processes. More specifically, $\calR$ will consist of a set of process \emph{families} $(R_t)$, each potentially parameterized by a sequence $(\lambda_t)$. To make things concrete, consider the process family defined by
	\begin{equation}
		\label{eq:example-process}
		\begin{gathered}
		R_0^\lambda=0,\ifarxiv\qquad\else\\\fi R_t^\lambda = R_t(\lambda_1^t) = C\exp\left\{\sum_{i=1}^t \lambda_i X_i - \frac{1}{8}\sum_{i=1}^t \lambda_i^2\right\},\quad t\geq1,
		\end{gathered}
	\end{equation}
	where we use $\lambda_1^t$ as shorthand for $(\lambda_1,\dots,\lambda_t)$. When opting into the contract, the agent would choose a specific reward process $(R_t^\lambda)$ but specify only $\lambda_1$ initially. Then, before round $t$, the agent chooses $\lambda_t$ based on past observations. This example was not pulled from whole cloth: the underlying exponential process, initialized at one, is a nonnegative supermartingale for iid data with mean 0 bounded in $[-1/2,1/2]$. (More generally, for $\lambda_i\geq 0$, it is a nonnegative supermartingale whenever the data are iid, have nonpositive mean and are $1/2$-sub-Gaussian, e.g., $X_i\sim N(\theta,1/4)$ with $\theta\leq 0$.) As we'll see below, nonnegative supermartingales result in null-deterring contracts.

	We will consider two reward schedules: deferred and incremental. Under the deferred schedule, the agent receives payment only when the experiment ends. Under the incremental schedule, the agent receives rewards along the way, with each period’s reward becoming immediately available. Incremental rewards are natural when the rewards are themselves immediately actionable. A pharmaceutical company may want to take their drug to a corner of the market, or begin profiting, based on the evidence thus far.
    Deferred rewards arise in settings where evidence is accumulated sequentially, but the relevant transfer or decision is made only once, at the end. Examples include regulatory approval, prize competitions, grant funding, or venture financing. In such settings, the agent may gather evidence over time, while the eventual reward is determined only when the data gathering stops.

	The reward schedule also restricts the reward process. Under the incremental schedule, previously disbursed rewards cannot be clawed back, so the cumulative reward should be nondecreasing over time. Under the deferred schedule, only the terminal reward is given. Thus, the total reward may fluctuate up and down according to the evidence.

    \ifarxiv
	With the example of the reward process above and the two reward schedules in mind, let us define a sequential statistical contract.
    \fi

	\begin{tcolorbox}[
		colback=blue!8!white,
		colframe=white,
		boxrule=0pt,
		sharp corners,
		left=6pt,
		right=6pt,
		top=6pt,
		bottom=6pt,
		breakable=true
		]
		\textbf{Sequential statistical contract} $\calC = (C, \calR, \sigma)$
		\begin{enumerate}[nosep,topsep=0.2cm]
			\item The principal \ifarxiv publicly \fi  announces a cost $C>0$, a menu of reward processes $\calR$, a stopping time $\sigma\geq1$ (which may be infinite), and the type of reward schedule.
			\item If the agent pays $C$ to opt-in, they choose a reward process $(R_t^\lambda)_{t\geq 0}$ from $\calR$. Set $R_0 = 0$.
			\item For $t=1,2,\dots$
			\begin{itemize}[nosep]
				\item The agent chooses parameter(s) $\lambda_t$ if applicable
				\item A trial is run producing data $X_t \sim P_\theta$, where $\theta$ is the agent's type
				\item The total reward $R_t = R_t(X_1^t,\lambda_1^t)\geq 0$ is calculated. On the incremental reward schedule, the agent is awarded $R_t - R_{t-1}$, ensured to be nonnegative.
                \item If $\sigma=t$ the contract is ended. On the deferred schedule, the agent is awarded $R_t$.
				\item The agent decides whether to stop or continue.
				\item If they stop, on the deferred schedule, they are awarded $R_t$.
			\end{itemize}
		\end{enumerate}
	\end{tcolorbox}

	Several remarks are in order. First, in addition to allowing the agent to stop the contract, we also allow the principal to commit ex ante to a stopping rule. This is useful in a variety of settings. For example, the principal may face a finite budget, a fixed review horizon, or some other operational constraint that limits how long the experiment can continue. Of course, the stopping rule need not be so explicit; it may depend on the growth rate of some utility function, for instance.

	Second, we do not impose per-period costs on the agent. That is, once the agent has paid the entry cost $C$, they do not make additional payments each round in order to generate further data. We interpret $C$ as the upfront cost of mounting the research program: setting up the experiment, building infrastructure, hiring personnel, obtaining approval, and incurring associated opportunity costs.  This differs from the sequential model of \citet{bates2022principal}; see Appendix~\ref{app:bates-comparison}.

	More broadly, the costs of continuation need not be symmetric across the two parties. For instance, once the FDA has authorized a clinical trial, their  marginal cost of allowing the study to continue may be quite low. But the agent may still face substantial costs from administering the treatment and monitoring patients.

	\section{Null Deterrence in Sequential Contracts}
	\label{sec:contracts}

	Following the intuition of \citet{bates2022principal} in the single-shot setting, the principal would like to design a menu $\calR$ such that, regardless of the agent's strategy (i.e., their choice of $R\in\calR$, their optimization of $(\lambda_t)$ if applicable, and when they choose to stop), the expected reward allocated to a null agent is at most $C$. In the single-shot setting, null deterrence (ND) is the condition that $\E_\theta[R(X)]\leq C$ for all $\theta\in\Theta_0$. We extend this to the sequential setting as follows.

	\begin{definition}
		\label{def:null-deterrence}
		The sequential contract $(C,\calR,\sigma)$ is \emph{null-deterring (ND)} if
        \ifarxiv
		\begin{equation*}
			\label{eq:nd}
			\text{for all $(R_t)\in\calR$ and all $\theta\in\Theta_0$, $\E_{\theta}[R_{\tau\wedge \sigma}]\leq C$ for any stopping time $\tau\geq1$.}  \tag{ND}
		\end{equation*}
        \else
        for all $(R_t)\in\calR$ and all $\theta\in\Theta_0$, $\E_{\theta}[R_{\tau\wedge \sigma}]\leq C$ for any stopping time $\tau\geq1$.\fi
	\end{definition}
	We may sometimes discuss null deterrence not of an entire contract itself, but of a single process $R$. In particular, if $R$ satisfies $\E_{\theta}[R_{\tau\wedge \sigma}]\leq C$ for all $\theta\in \Theta_0$ and all stopping times $\tau\geq1$,\footnote{Stopping times may be infinite. On $\{\tau\wedge\sigma=\infty\}$ we set $R_\infty:=0$ on the deferred schedule and $R_\infty:=\lim_{t\to\infty}R_t$ on the incremental schedule. The limit exists because rewards are nondecreasing.} we will say that it is ND.
	Importantly, null deterrence does not require the principal to know the agent's utility function. The condition holds under every agent strategy and it is thus an asymmetric guarantee, prioritizing the principal over the agent.

	On the deferred reward schedule, \ifarxiv\eqref{eq:nd}\else Definition~\ref{def:null-deterrence}\fi\ should remind us of the definition of an e-process. Indeed, we will show below that all ND contracts on the deferred reward schedule are characterized by e-processes.

	Things are not quite as simple on the incremental reward schedule, however. Many, if not most, e-processes are not almost surely increasing, and thus do not satisfy $R_t-R_{t-1}\geq 0$ for all $t\geq 1$.
	Instead, we can construct ND contracts by using the concept of an \emph{adjuster}~\citep{shafer2011test,choe2026combining,dawid2011probability,dawid2011insuring}, which is a nonnegative, nondecreasing, right-continuous function $A:[1,\infty]\to[0,\infty]$ with $A(\infty)=\lim_{y\to\infty}A(y)$, satisfying
	\begin{equation}
    \label{eq:adjuster-def}
		\int_1^\infty \frac{A(y)}{y^2} \d y\leq 1.
	\end{equation}
    The adjuster is \emph{admissible} if~\eqref{eq:adjuster-def} holds with equality and $A(\infty)=\infty$~\citep{shafer2011test}.
	Adjusters have their roots in the game-theoretic interpretation of e-processes, where they are viewed as the wealth process of a statistician playing a game against nature~\citep{ramdas2023game}. In such a setting, it is sometimes convenient to ensure that the wealth doesn't decrease over time, which is what adjusters were developed to accomplish.

	More formally, an adjuster satisfies the following guarantee: For any test supermartingale $(M_t)$ (meaning a nonnegative supermartingale with initial value 1) there exists a test supermartingale $(Y_t)$ such that $A(M_t^*)\leq Y_t$ almost surely, where $M_t^* = \max_{0\leq s\leq t} M_s$~\citep[Section 7]{shafer2011test}. This, together with the optional stopping theorem, implies that $A(M_t^*)$ is an e-process, a fact which will undergird our results.

    \begin{figure}[t]
		\centering
        \ifarxiv
		\includegraphics[width=0.92\textwidth]{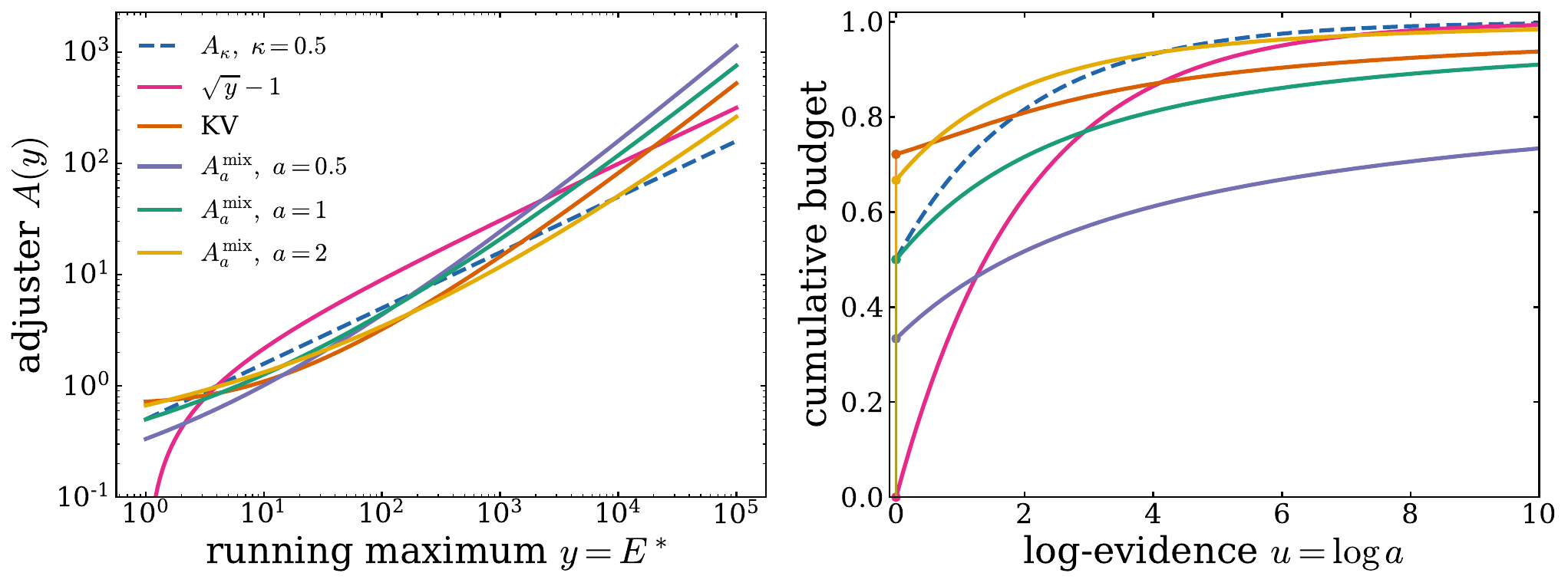}
        \else
        \includegraphics[width=0.45\textwidth]{adjuster_growth_stacked.pdf}
        \fi
\caption{Depictions of the Shafer adjuster $A_\kappa(y)=\kappa y^{1-\kappa}$ with $\kappa=0.5$, the square-root adjuster $\sqrt y-1$~\citep{choe2026combining}, the KV adjuster $y^2\log2/((1+y)\log^2(1+y))$~\citep{koolen2014buy}, and the $\mathrm{Beta}(a,1)$ mixtures $A_a^{\mix}$ for $a\in\{0.5,1,2\}$. \ifarxiv\emph{Left:}\else\emph{Top:}\fi\space the reward as a function of running maximum evidence. \ifarxiv\emph{Right:}\else\emph{Bottom:}\fi\space the share of the null-payout budget spent on payments at evidence levels up to $e^u$, namely $A(1)+\int_{(1,e^u]}\d A(a)/a$, against log-evidence $u$. Dots at zero show the unconditional first-round payment $CA(1)$ as a fraction of $C$, and the remaining budget funds later milestones.}
		\label{fig:adjusters}
	\end{figure}

	With the concept of adjusters in hand, we can state the following result regarding ND statistical contracts.

	\begin{proposition}
		\label{prop:null-deterrence}
		Let $\calC = (C,\calR,\sigma)$ be a sequential statistical contract.
		For the deferred reward schedule, $\calC$ is ND if and only if $(R_{t\wedge \sigma}/C)$ is an e-process for $\Theta_0$ for all $(R_t)\in\calR$.
		For the incremental reward schedule, suppose every $(R_t)\in\calR$ is nondecreasing. Then $\calC$ is ND if and only if $\E_\theta[R_\infty^\sigma]\leq C$ for all $\theta\in\Theta_0$ and all $(R_t)\in\calR$, where $R_\infty^\sigma = \lim_{t\to\infty}R_{t\wedge\sigma}$. In particular, $\calC$ is ND if $(R_t/C)$ is an e-process for $\Theta_0$ for every $(R_t)\in\calR$.
		Further, given an e-process $(E_t)_{t\geq 1}$ for $\Theta_0$, $(C A(E_t^*))_{t\geq 1}$ is ND for any adjuster $A$, where $E_t^* = 1\vee\max_{1\leq k\leq t} E_k$.
	\end{proposition}

Faced with Proposition~\ref{prop:null-deterrence}, a natural question is whether all increasing e-processes can be represented via adjusters. This turns out to be false.  Appendix~\ref{app:adjuster-counterexample} gives a bounded nondecreasing e-process that cannot be represented, or even dominated, by an adjusted running maximum.

Nevertheless, adjusted e-processes will play an important role in our analysis in Section~\ref{sec:agent-best-response}, so let us linger on their details here. To foster some intuition,  consider the following scheme. The principal fixes evidence levels $1<a_1<\cdots<a_K$, which we call \emph{milestones}, and promises a \emph{milestone payment} $r_j\geq0$ the first time the e-process $(E_t)$ (i.e., the evidence) reaches $a_j$. The cumulative rewards are then
\begin{equation}
\label{eq:milestone-contract}
R_t=\sum_{j=1}^K r_j\ind{E_t^*\geq a_j} \text{~ such that ~}
\sum_{j=1}^K\frac{r_j}{a_j}\leq C,
\end{equation}
where $E_t^*$ is the running maximum. When the agent is null, the probability that the evidence ever reaches $a_j$ is at most $1/a_j$ by Ville's inequality~\citep{ville1939etude,ruf2023composite}, so a payment of $r_j$ at level $a_j$ costs at most $r_j/a_j$ in expectation. The constraint in~\eqref{eq:milestone-contract} thus caps the expected payout to a null agent at the entry cost $C$, which we call the \emph{null-payout budget}, and it is exactly~\eqref{eq:adjuster-def} for $A(y)=C^{-1}\sum_jr_j\ind{y\geq a_j}$. On this view, an adjuster is simply a choice of how to spend the null-payout budget across evidence levels, with milestone $a_j$ using the fraction $r_j/(Ca_j)$.

The same reading applies to any adjuster, not only those with discrete milestone payments. Since $A$ is nondecreasing and right-continuous,
\begin{equation}
\label{eq:adjuster-milestone-representation}
A(y)=A(1)+\int_{(1,\infty)}\ind{y\geq a}\d A(a),
\end{equation}
and Tonelli's theorem turns~\eqref{eq:adjuster-def} into the budget constraint
\begin{equation}
\int_1^\infty\frac{A(y)}{y^2}\d y=A(1)+\int_{(1,\infty)}\frac{\d A(a)}{a}\leq1.
\end{equation}
The contract therefore pays a baseline $CA(1)$ at the end of the first round, whatever the data, and pays $C\d A(a)$ the first time the evidence reaches $a>1$. The baseline is reached with probability one, so it costs its full value, whereas a payment at level $a$ costs only the fraction $1/a$ of its value.

	\citet{shafer2011test} suggested the class of admissible adjusters $A_\kappa(y) = \kappa y^{1-\kappa}$ for $\kappa\in(0,1)$. Other examples include $A(y) = \sqrt{y}-1$~\citep{choe2026combining} and  $A(y) = \frac{y^2\log 2}{(1 + y)\log^2(1 + y)}$~\citep{koolen2014buy}.  We can also consider mixing over $A_\kappa$:
	\[A_a^\mix(y) = \int_0^1 \kappa y^{1-\kappa} a\kappa^{a-1}\d\kappa = ay\int_0^1 \kappa^a y^{-\kappa}\d\kappa.\]
	The uniform mixture $a=1$ recovers the adjuster used by \citet{choe2026combining}.

	Figure~\ref{fig:adjusters} compares the growth rates of these adjusters. Although all are admissible, they allocate their null-payout budget differently \ifarxiv (right panel)\else (bottom panel)\fi. The baseline reward matters. In particular, $A_{1/2}(1)=1/2$ and $A^{\mix}_1(1)=1/2$, so these adjusters return half of the entry cost in round one, while the KV adjuster returns $1/(2\log2)\approx72\%$ and the square-root adjuster has no baseline payment. The principal's choice of adjuster thus encodes a preference over where on the evidence scale to reward the agent.

    \ifarxiv
    \begin{remark}
    Definition~\ref{def:null-deterrence} was formulated linearly in terms of the reward and cost. One could instead require $\E_\theta[g(R_{\tau\wedge\sigma})]\leq g(C)$ for a nonnegative measurable function $g$ with $g(C)>0$. The e-process characterization in Proposition~\ref{prop:null-deterrence} then applies to $g(R_{t\wedge\sigma})/g(C)$ for $t\geq1$, extended by zero at time zero.
    \end{remark}
    \fi

\section{The Agent's Strategy}
\label{sec:agent-best-response}

The discussion on null-deterring contracts in the preceding section focuses on designing useful contracts from the principal's perspective. But what about the agent? Faced with a null-deterring contract, what is the agent's best response?

To answer this, we need to endow the agent with a utility function. Once such a function is introduced, the agent's optimal strategy can be handled using machinery from stochastic control theory~\citep{puterman1990markov,shiryaev2008optimal}. In particular, it can be written in terms of a Bellman equation (or the Snell envelope, if the parameter policy is fixed in advance). We give the details in Appendix~\ref{sec:full-knowledge}. But such solutions are not closed-form, not necessarily computationally tractable, and require that the agent possess full knowledge of their data-generating distribution $P_\theta$. Therefore, in this section, we are interested in strategies that are more transparent, easier to interpret, and lend themselves to approximations that do not require full knowledge of the distribution.

The Bellman and Snell approaches given in the appendix handle general utility functions and e-processes, but our approximations require making a few simplifying assumptions. For one, we will study specific utility functions for the agent, one for the incremental reward schedule and one for the deferred schedule.
For a fixed \emph{discount factor} $0<\delta<1$ and $t\geq1$, define
\begin{equation*}
\begin{gathered}
G_t^\inc:=\sum_{s=1}^t\delta^s(R_s-R_{s-1}),
\ifarxiv\text{~ and ~}\else\\\fi
G_t^\dfrd:=\delta^t\log(R_t).
\end{gathered}
\end{equation*}
The superscripts refer to the reward schedule. Under the deferred schedule the agent is paid once, so only the final payout matters; under the incremental schedule every interim payment counts. Both objectives are common in financial models~\citep[see, e.g.,][]{asmussen1997controlled,karatzas2000utility}.

The two reward schedules (and their corresponding utility functions) affect the agent's decision problem. Since the reward process must be nondecreasing on the incremental schedule, once the agent opts into the contract they should never exit. Therefore, the agent's decision on the incremental schedule is how to choose the parameters $(\lambda_t)$ (if any) to optimize the reward process. On the deferred schedule, meanwhile, the agent is concerned both with optimizing these parameters and also with potentially stopping and exiting the contract. Indeed, there's no guarantee that $(G_t^\dfrd)$ is nondecreasing in $t$.

We also assume throughout the rest of this section that the principal imposes no stopping time ($\sigma=\infty$) and that the underlying e-process takes the form of a product:
\begin{equation}
\label{eq:product-eprocess}
\begin{gathered}
E_t^\pi=\prod_{i=1}^t f_{\lambda_i}(X_i),\qquad E_0^\pi=1,
\ifarxiv\qquad\else\\\fi W_t^\pi=\log E_t^\pi,
\end{gathered}
\end{equation}
where $f_\lambda$ is positive and the superscript $\pi$ is the strategy determining the parameters $\lambda_i$ and, if applicable, the stopping time.  For $t\geq1$, the rewards under the deferred schedule will be $R_t^\pi=CE_t^\pi$ and those under the incremental schedule will be $R_t^\pi=CA((E^\pi)_t^*)$ for some adjuster $A$.

Let us summarize our path forward.

\begin{enumerate}
\item First we identify a parameter $\lambda_\cgf$ defined by
\begin{equation}
\begin{gathered}
\lambda_\cgf \in \argmax_{\lambda\in\Lambda} \E_\theta[f_\lambda(X)^{\rho^*}] \ifarxiv\text{ ~ where ~}\else\text{ where}\\\fi \sup_{\lambda\in\Lambda} \E_\theta[f_\lambda(X)^{\rho^*}] = \frac{1}{\delta},
\end{gathered}
\end{equation}
where $\Lambda$ is the set of values $\lambda$ can take. We call $\lambda_\cgf$ the \emph{CGF-optimal} parameter, so named because it can be defined by the CGF of the log-increment $\log f_\lambda(X)$, as we'll see below. Under both reward schedules, we bound the gap between the agent's utility under $\lambda_\cgf$ and the optimal adaptive strategy as a function of $\rho^*$. In particular, the gap is small when $\rho^*h$ is small, where $h$ bounds the log-increments.
\item We then introduce the \emph{growth-optimal parameter} $\lambda_\gro$, a first-order approximation to $\lambda_\cgf$, defined by
\begin{equation}
\lambda_\gro\in \argmax_{\lambda \in\Lambda} \E_\theta[\log f_\lambda(X)],
\end{equation}
which can be related to $\lambda_\cgf$ via $\log \E_\theta[f_\lambda(X)^{\rho}] = \rho \E_\theta[\log f_\lambda(X)] + O(\rho^2)$.  The GRO objective is easier to estimate from data by an empirical average and thus motivates several plug-in policies that can be instantiated even when the agent does not have full knowledge of their distribution. These plug-ins are presented in Section~\ref{sec:drift-policies} and the strategies are compared experimentally in Section~\ref{sec:examples}.
\end{enumerate}
We begin by introducing the CGF-optimal parameter $\lambda_\cgf$.

\subsection{The CGF-optimal rule}
\label{sec:discounted-evidence}

For a fixed parameter $\lambda$, define the iid log-increments and their drift  by
\begin{equation}
\label{eq:log-increments}
\Delta_i(\lambda):=\log f_\lambda(X_i),\qquad
\mu(\lambda):=\E_\theta[\Delta_1(\lambda)].
\end{equation}
Recall that $W_t^\pi$ is the logarithm of the e-process in~\eqref{eq:product-eprocess}.
For a threshold $\kappa\geq0$, write
\begin{equation}
\label{eq:discounted-crossing}
\begin{gathered}
T_\pi(\kappa)=\inf\{t\geq0:W_t^\pi\geq\kappa\},
\ifarxiv\qquad\else\\\fi H_\pi(\kappa)=\E_{\theta,\pi}[\delta^{T_\pi(\kappa)}],
\end{gathered}
\end{equation}
with $\delta^\infty=0$. In words, $T_\pi(\kappa)$ is the first time that the log-evidence $W_t^\pi$ reaches $\kappa$, and $H_\pi(\kappa)$ is the expected discount at that time. For a strategy that plays a constant parameter $\lambda$, we write $T_\lambda(\kappa)$ and $H_\lambda(\kappa)$.

The reader may be wondering why we defined $T_\pi$. Why are we studying threshold rules? First, they are natural under the incremental schedule when the reward process is an adjusted e-process, which pays $C\d A(a)$ exactly when the evidence first reaches $a$ (Section~\ref{sec:contracts}). Second, under the deferred schedule with discounted log utility, a reward-threshold rule is optimal at each fixed parameter with integrable log-increments (Appendix~\ref{app:fixed-stopping-details}).

Fix $\lambda$ and define the CGF
\begin{equation}
\psi_\lambda(s)=\log\E_\theta[e^{s\Delta_1(\lambda)}].
\end{equation}
Consider what happens as the strategy which plays $\lambda$ tries to reach a log-evidence of $\kappa$. Each round changes the log-evidence by $\Delta_t(\lambda)$ and multiplies the eventual discount by $\delta$. Since the increments are iid, conditioning on the first one gives the recurrence
$H_\lambda(\kappa) = \delta \E_\theta[H_\lambda(\kappa - \Delta_1(\lambda))]$ for $\kappa>0$, with $H_\lambda(\kappa)=1$ for $\kappa\leq0$. If, for the moment, we posit that the solution to this recursive equation looks like $H_\lambda(\kappa) = \exp(-s\kappa)$ for some fixed $s$, then taking logs of both sides and rearranging gives $\log(1/\delta) = \log \E_\theta[\exp(s\Delta_1)] = \psi_\lambda(s)$. This motivates the definitions
\begin{equation}
\label{eq:cgf-exponents}
\rho_\lambda:=\sup\{s\geq0:\psi_\lambda(s)\leq\log(1/\delta)\},
\qquad \rho^*:=\inf_{\lambda\in\Lambda}\rho_\lambda.
\end{equation}
For well-behaved CGFs, we have $\psi_\lambda(\rho_\lambda) = \log(1/\delta)$. Then $\rho^*$ is simply the smallest across all such roots and, when $0<\rho^*<\infty$, it satisfies $\delta\E_\theta[f_\lambda(X)^{\rho^*}]\leq1$ for every $\lambda\in\Lambda$.\footnote{Each $\psi_\lambda$ is convex and lower semicontinuous with $\psi_\lambda(0)=0<\log(1/\delta)$, so $\{s\geq0:\psi_\lambda(s)\leq\log(1/\delta)\}$ is a closed interval containing zero. Since $\rho^*\leq\rho_\lambda$, it follows that $\psi_\lambda(\rho^*)\leq\log(1/\delta)$.}

The approximation $H_\lambda(\kappa)\approx e^{-\rho_\lambda\kappa}$ suggests minimizing $\rho_\lambda$. For $0<\rho^*<\infty$, we call a parameter $\lambda_\cgf$ \emph{CGF-optimal} if it attains the infimum and satisfies the recurrence relation with equality:
\begin{equation}
\label{eq:cgf-parameter}
\lambda_\cgf \in \argmin_{\lambda\in\Lambda} \rho_\lambda \text{~ and ~} \delta \E_\theta[f_{\lambda_\cgf}(X)^{\rho^*}]=1.
\end{equation}
Equivalently, the CGF-optimal parameter satisfies $\psi_{\lambda_\cgf}(\rho^*)=\log(1/\delta)$.
Such a parameter maximizes $\E_\theta[f_\lambda(X)^{\rho^*}]$ over $\lambda\in\Lambda$.
Only the lower bounds below require that the minimum in~\eqref{eq:cgf-parameter} is achieved. The upper bounds hold regardless.

Intriguingly, the exponent $\rho^*$ has recently appeared in another context: it is the optimal exponent $\eta_\star$ of the \emph{exponential-decay-optimal} (EDO) criterion of \citet[Section~7, equation~(4)]{clerico2026time}. The same exponent appears because both problems discount the time needed to reach a given level of evidence. Proposition~\ref{prop:crossing-comparison} below is thus similar to their Proposition~3.

The motivation for $\lambda_\cgf$ and $\rho^*$ rests on the approximation $H_\lambda(\kappa)\approx e^{-\rho_\lambda\kappa}$. We might wonder how tight this is. The following result shows that $H_\pi(\kappa)\leq e^{-\rho^*\kappa}$ for every strategy $\pi$, and that a fixed parameter nearly attains this bound when the \emph{overshoot} (the amount by which the log-evidence exceeds $\kappa$ at the crossing time) is small.

\begin{proposition}
\label{prop:crossing-comparison}
Suppose the observations are iid under $P_\theta$, and let $s\in(0,\infty)$ satisfy $
\delta\E_\theta[f_\lambda(X)^s]\leq1$ for every $\lambda\in\Lambda$.
Then, for every predictable parameter strategy $\pi$, the process $(\delta^t(E_t^\pi)^s)$ is a nonnegative supermartingale under $P_\theta$, and $H_\pi(\kappa)\leq \exp\{-s\kappa\}$ for all $\kappa\geq 0$.
Further, if a fixed parameter $\lambda_0\in\Lambda$ satisfies $\delta\E_\theta[f_{\lambda_0}(X)^s]=1$ and $\Delta_1(\lambda_0)\leq h$ almost surely for some $h>0$, then, for every $\kappa\geq0$,
\begin{equation}
\label{eq:crossing-comparison}
\exp\{-s(\kappa+h)\}\leq H_{\lambda_0}(\kappa) \leq\sup_\pi H_\pi(\kappa)\leq \exp\{-s\kappa\}.
\end{equation}
\end{proposition}
Thus, taking $s=\rho^*$ and $\lambda_0=\lambda_\cgf$, whenever the stated assumptions hold, gives $H_{\lambda_\cgf}(\kappa)\geq e^{-\rho^*h}\sup_\pi H_\pi(\kappa)$.
That is, the fixed CGF rule achieves at least the fraction $e^{-\rho^*h}$ of the optimal discounted value under any adaptive strategy. The proof is in Appendix~\ref{app:proofs}.

Next we study the properties of  the CGF-optimal rule under each reward schedule. Before we do, however, let us emphasize that the CGF-optimal rule still requires essentially full distributional knowledge from the agent, since a CGF that is finite near zero determines the distribution. Thus, it requires about as much information as the Bellman equation. However, it is significantly more transparent and, via results such as Proposition~\ref{prop:crossing-comparison}, allows us to analyze how close it is to the optimal strategy. Moreover, as we'll see in Section~\ref{sec:drift-policies}, it motivates various plug-in strategies which can be implemented under limited (or no) distributional knowledge.

\subsection{\texorpdfstring{$\lambda_\cgf$}{λ\_CGF} under both reward schedules}
\label{sec:cgf-schedules}

We begin with the incremental reward schedule.
Fix an adjuster $A$ and cumulative rewards $R_t^\pi=CA((E^\pi)_t^*)$ for $t\geq1$.  The agent chooses predictable parameters to maximize
\begin{equation}
\label{eq:incremental-adjuster-value}
\begin{aligned}
J_{\mathrm{inc}}(\pi)
&:=\E_{\theta,\pi}\bigg[\sum_{t\geq1}\delta^t(R_t^\pi-R_{t-1}^\pi)\bigg]\aaaibreak=\delta CA(1)+C\int_{(1,\infty)}H_\pi(\log a)\d A(a),
\end{aligned}
\end{equation}
where the second equality follows from the adjuster representation~\eqref{eq:adjuster-milestone-representation} in Section~\ref{sec:contracts}.
We can integrate the crossing bounds in Proposition~\ref{prop:crossing-comparison}, taking $s=\rho^*$ and $\kappa=\log a$, to obtain bounds on the agent's utility as follows.

\begin{proposition}
\label{prop:incremental-guarantee}
Suppose the observations are iid under $P_\theta$ and $0<\rho^*<\infty$. Let $A$ be an adjuster and consider rewards $R_t^\pi=CA((E^\pi)_t^*)$ for $t\geq1$ and $\sigma=\infty$. Define
\[
\begin{gathered}
U_A:=\delta CA(1)+C\int_{(1,\infty)}a^{-\rho^*}\,\d A(a),
\ifarxiv\qquad\else\\\fi V_A^\star:=\sup_\pi J_{\mathrm{inc}}(\pi),
\end{gathered}
\]
and suppose $U_A<\infty$. Then $V_A^\star\leq U_A$. If, in addition, there exists a parameter $\lambda_\cgf\in\Lambda$ satisfying $\delta\E_\theta[f_{\lambda_\cgf}(X)^{\rho^*}]=1$ and $\Delta_1(\lambda_\cgf)\leq h$ almost surely for some $h>0$, then
\begin{equation}
\label{eq:incremental-guarantee}
e^{-\rho^*h}U_A
\leq J_{\mathrm{inc}}(\lambda_\cgf)
\leq V_A^\star
\leq U_A.
\end{equation}
\end{proposition}

\begin{remark}
The upper bound also tells us when positive expected discounted profit is impossible. If $\rho^*\geq1$, then $a^{-\rho^*}\leq a^{-1}$ for $a\geq1$, so the adjuster budget gives
\[
V_A^\star\leq U_A\leq\delta CA(1)+C\int_{(1,\infty)}a^{-1}\,\d A(a)
\leq C.
\]
Thus $\rho^*<1$ is necessary for the agent's expected discounted receipts to exceed the entry cost $C$.
\end{remark}

Next let us consider the deferred schedule. Here, the agent chooses both a parameter strategy and when to collect the reward. Stopping times $\tau\geq1$ may be infinite, in which case the agent never collects and we set $G_\infty^\dfrd:=0$. For any fixed parameter $\lambda$ with integrable log-increments, a threshold rule attains the optimal stopping value (Appendix~\ref{app:fixed-stopping-details}).
We thus study the utility $G^\dfrd$ under the threshold rules $T_\lambda(\kappa)$.
For $\kappa>0$, define
\begin{equation}
\label{eq:threshold-value-exact}
J_\lambda(\kappa):=\E_\theta\left[G_{T_\lambda(\kappa)}^\dfrd\right] = \E_\theta\left[\delta^{T_\lambda(\kappa)} \log R_{T_\lambda(\kappa)}(\lambda)\right].
\end{equation}
At the time $T=T_\lambda(\kappa)$, we have $W_T=\log E_T\geq\kappa$ and $\log R_T=\log C+W_T$. Ignoring overshoot, so that $W_T\approx\kappa$, gives the following approximation to the expected discounted utility:
\begin{equation}
\label{eq:cgf-surrogate}
\begin{gathered}
J_\lambda(\kappa)\approx\widehat J_\lambda(\kappa)
:=e^{-\rho_\lambda\kappa}(\log C+\kappa),
\end{gathered}
\end{equation}
where we're using that $H_\lambda(\kappa) \approx \exp(-\rho_\lambda\kappa)$.
When $\log C<1/\rho_\lambda$, maximizing this approximation over $\kappa>0$ gives
\begin{equation}
\label{eq:cgf-maximized-value}
\kappa_\lambda^\cgf=1/\rho_\lambda-\log C,\qquad
\widehat J_\lambda(\kappa_\lambda^\cgf)=C^{\rho_\lambda}/(e\rho_\lambda).
\end{equation}
In this regime, the maximized value is a decreasing function of $\rho_\lambda$, hence $\lambda_\cgf$, when it exists, also optimizes this approximation. Thus, the \emph{CGF-optimal policy} under the deferred schedule uses $\lambda_t\equiv\lambda_\cgf$ and the stopping rule
\begin{equation}
\label{eq:cgf-policy}
\tau_\cgf=\inf\{t\geq1:R_t(\lambda_\cgf)\geq e^{1/\rho^*}\}.
\end{equation}
For any other fixed parameter $\lambda$ with $0<\rho_\lambda<\infty$, the corresponding reward threshold is $R_\cgf(\lambda)=e^{1/\rho_\lambda}$. As in the case of incremental rewards, we can use properties of $\rho^*$ to give optimality guarantees for the CGF-optimal parameter and  stopping rule.

\begin{proposition}
\label{prop:cgf-guarantee}
Suppose the observations are iid under $P_\theta$ and $0<\rho^*<\infty$.
Consider deferred rewards $R_t^\pi=CE_t^\pi$ for $t\geq1$, with $C>0$. Let $V_\theta^\star(C)$ be the supremum of expected utility over predictable parameter strategies and all stopping times $\tau\geq1$, and define
\begin{equation}
\label{eq:cgf-upper-general-c}
U_{\dfrd}(C):=
\begin{cases}
C^{\rho^*}/(e\rho^*), & \log C<1/\rho^*,\\
\log C, & \log C\geq1/\rho^*.
\end{cases}
\end{equation}
Then $V_\theta^\star(C)\leq U_{\dfrd}(C)$. If $\log C<1/\rho^*$, suppose also that there exists a parameter $\lambda_\cgf\in\Lambda$ satisfying $\delta\E_\theta[f_{\lambda_\cgf}(X)^{\rho^*}]=1$ and $\Delta_1(\lambda_\cgf)\leq h$ almost surely for some $h>0$.
Set $\kappa=1/\rho^*-\log C$ and $q=(1+\rho^*h)e^{-\rho^*h}$. Then
\begin{equation}
\label{eq:deferred-cgf-guarantee}
qU_{\dfrd}(C)
\leq J_{\lambda_\cgf}(\kappa)
\leq V_\theta^\star(C)
\leq U_{\dfrd}(C).
\end{equation}
\end{proposition}

Thus the CGF-optimal policy attains at least the fraction $q$ of the optimal expected utility over all predictable strategies and stopping times. Appendix~\ref{app:proofs} gives a corresponding guarantee for unbounded log-increments in terms of the overshoot's second moment.

\subsection{Approximate strategies}
\label{sec:drift-policies}

Let us now suppose that the agent does not know their full distribution. In this case they can try to approximate the CGF-optimal strategy via a first-order Taylor expansion. For fixed $\lambda$ with $\mu(\lambda)>0$ and a CGF finite near zero, $\psi_\lambda(s)=s\mu(\lambda)+o(s)$ gives $\rho_\lambda\approx \log(1/\delta)/\mu(\lambda)$ as $\delta\to 1$. (Discount factors close to one, which describe patient agents, are the natural regime.) Therefore,  to minimize $\rho_\lambda$ we want to maximize $\mu(\lambda)$, which suggests choosing
\begin{equation}
\lambda_\gro \in \argmax_{\lambda\in\Lambda} \mu(\lambda).
\end{equation}
Under the incremental schedule the agent never stops, so they use $\lambda_\gro$ throughout. Under deferred rewards, substituting $1/\rho_\lambda\approx\mu(\lambda)/\log(1/\delta)$ and following~\eqref{eq:cgf-policy} gives the threshold
\begin{equation}
\label{eq:drift-reward-threshold}
R_\drift(\lambda)=\exp\{\mu(\lambda)/\log(1/\delta)\}.
\end{equation}
The agent collects at the first round $t\geq1$ at which $R_t(\lambda)$ reaches this threshold.

Of course, these choices require knowing the drift. But there are many distributions for which it is reasonable to know the drift, even when full distributional knowledge is unavailable. For the exponential reward~\eqref{eq:example-process}, for instance, $\mu(\lambda)=\lambda\theta-\lambda^2/8$ depends only on the observation mean $\theta=\E_\theta[X]$. The same holds for ``sub-$\psi$'' processes~\citep{howard2020time} with a deterministic variance process, which cover a wide range of e-processes.

Further, when the drift is unknown, the agent can estimate it with the plug-in estimator
\begin{equation}
\muhat_t(\lambda)=\frac1t\sum_{i=1}^t\log f_\lambda(X_i),
\end{equation}
and bet an approximate maximizer $\widehat\lambda_t\in\Lambda$ of $\muhat_t$ in the next round.  Since the contract requires $\lambda_t$ to be chosen before $X_t$ is observed, the agent starts from some $\lambda_1\in\Lambda$ and, after observing $X_t$, sets $\lambda_{t+1}=\widehat\lambda_t$. The resulting e-process is $E_t^\pi=f_{\lambda_1}(X_1)\prod_{i=2}^t f_{\widehat\lambda_{i-1}}(X_i)$. Such a strategy has been called the ``method of predictable plug-ins''~\citep{waudby2024estimating} in other contexts. Examples are given in Section~\ref{sec:examples}. Under the deferred reward schedule, we use the following stopping time alongside the plug-in:
\begin{equation}
\label{eq:plugin-threshold}
\begin{aligned}
&\tau_\plugin=\inf\Bigl\{t\geq1:\log R_t\geq\aaaibreak[\qquad]
\max\Bigl(\log C,\frac{\muhat_t(\widehat\lambda_t)}{\log(1/\delta)}\Bigr)\Bigr\}.
\end{aligned}
\end{equation}
The floor at $\log C$ keeps the agent from collecting less than the entry cost $C$. Next let us experimentally compare all the strategies we've discussed.

\section{Numerical Illustrations}
\label{sec:examples}

We compare the strategies under both reward schedules, setting $C=1$. Appendix~\ref{app:sims} provides the implementation details.

\subsection{Incremental schedule}
\label{sec:incremental-experiment}

We consider Bernoulli observations with mean $p=0.7$, null mean at most $m=0.5$, and increments  $f_\lambda(x)=1+\lambda(x-m)$ for $\lambda\in[0,1/m)$. The resulting ``betting-style'' e-process, $E_t = \prod_{i\leq t} f_\lambda(X_i)$ is a common one in the e-value literature~\citep{waudby2024estimating}. Since we are working under the incremental reward schedule, the reward process is $CA(E_t^*)$ for some adjuster $A$.

We are interested in understanding the relative performance of our strategies compared to the optimal  Bellman benchmark (Appendix~\ref{app:additional-results}) as well as the effect of different adjusters. We consider two adjusters with three discrete milestone payments (see Section~\ref{sec:contracts}), which differ in where on the evidence scale they spend their null-payout budget.
Milestone $a_j$ uses the fraction $w_j$ of the budget, so the reward is $r_j=w_ja_j$, with $(a_1,a_2,a_3)=(2,4,8)$. The \emph{low-milestone} adjuster uses weights $(0.8,0.1,0.1)$ and the \emph{high-milestone} adjuster uses $(0.1,0.1,0.8)$.

\begin{figure*}[t]
\centering
\ifarxiv
\includegraphics[width=0.8\textwidth]{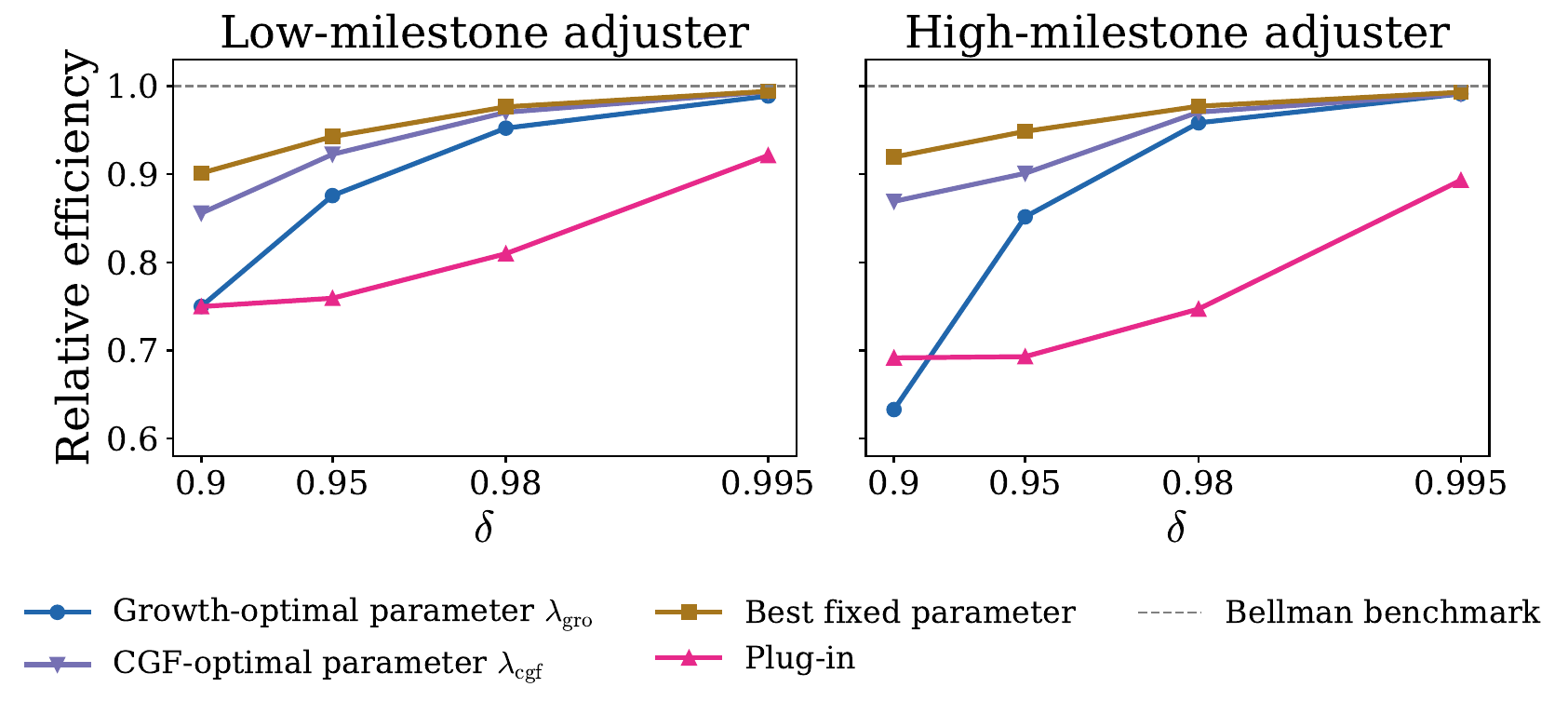}
\else
\includegraphics[width=0.7\textwidth]{incremental_milestones.pdf}
\fi
\caption{The utility of various strategies under the incremental reward schedule relative to the  Bellman benchmark~\eqref{eq:bellman}. The low- and high-milestone adjusters allocate $80\%$ of their null-payout budget (see Section~\ref{sec:contracts}) to the first and last milestone, respectively.
The plug-in is that of Section~\ref{sec:drift-policies} and uses the clipped sample mean. Standard errors of the ratios are below $0.0016$, so error bars are omitted. }
\label{fig:incremental-milestones}
\end{figure*}

Figure~\ref{fig:incremental-milestones} shows that all three fixed policies plotted attain at least $98\%$ of the Bellman value when $\delta=0.995$. As agents become less patient and $\delta$ decreases, the gap widens. In other words, with stronger discounting, adaptive parameter choice matters more, regardless of the adjuster.
At $\delta=0.9$ under the high-milestone adjuster, the growth-optimal rule earns $63.3\%$ of the Bellman value and the CGF-optimal rule $86.9\%$. We also compare both rules with the best fixed parameter, found numerically. Its gap to the CGF-optimal rule reflects the cost of overshoot in Proposition~\ref{prop:incremental-guarantee}.

Although the plug-in improves as the agent becomes more patient, it still falls short of the fixed policies at $\delta=0.995$. To see why, relate it to the growth-optimal rule. In the Bernoulli example, the plug-in uses $\widehat\lambda_t=(\widetilde p_t-m)^+/(m(1-m))$, with $\widetilde p_t=\min\{\bar X_t,0.99\}$, where $\bar X_t$ is the running mean. It starts at $\lambda_1=0.1$ and sets $\lambda_{t+1}=\widehat\lambda_t$. This is exactly $\lambda_\gro=(p-m)/(m(1-m))$ with $p$ replaced by $\widetilde p_t$, so the plug-in is the growth-optimal rule with an estimated parameter. For $\delta$ close to one, $\lambda_\gro$ is nearly optimal, $\widehat\lambda_t$ converges to it, and the plug-in's gap reflects the cost of learning. At $\delta=0.9$, however, $\lambda_\gro=0.8$ is far below the best fixed parameter of about $1.36$. The plug-in's early estimates are noisy, and after early successes they can be much larger than $\lambda_\gro$, moving the bet toward the better parameter. As a result, the plug-in matches the growth-optimal rule under the low-milestone adjuster and exceeds it under the high-milestone adjuster ($69.2\%$ versus $63.3\%$ of the Bellman value).

\subsection{Deferred schedule}
\label{sec:deferred-experiments}

\begin{figure*}[t]
\centering
\ifarxiv
\includegraphics[width=0.8\textwidth]{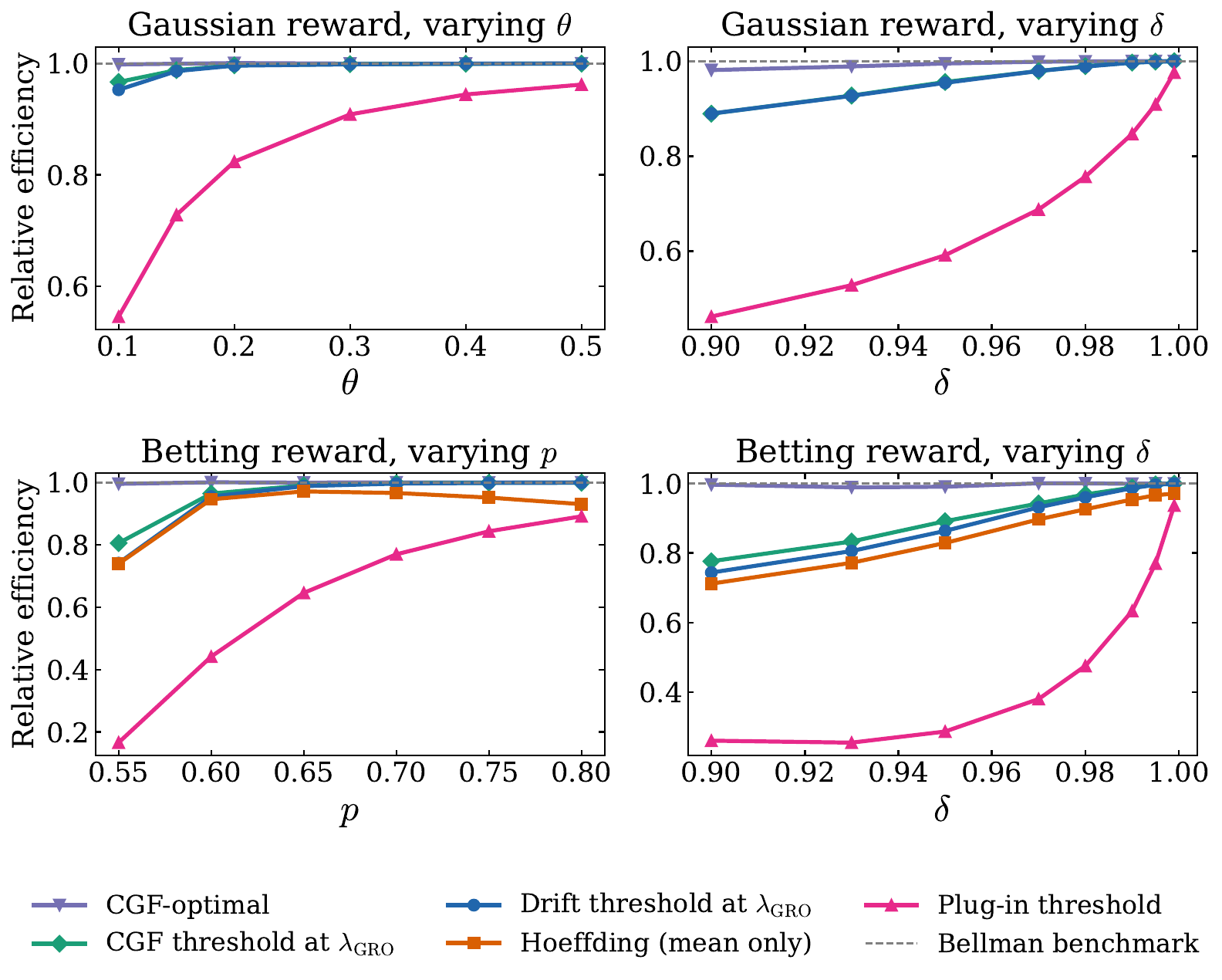}
\else
\includegraphics[width=\textwidth]{efficiency_row.pdf}
\fi
\caption{Deferred utility relative to the Bellman benchmark~\eqref{eq:bellman}. Sweeps over $\theta$ and $p$ use $\delta=0.995$, and sweeps over $\delta$ use $\theta=0.3$ (Gaussian) and $p=0.7$ (Bernoulli), with $m=0.5$. The CGF-optimal policy~\eqref{eq:cgf-policy} uses $\lambda_\cgf$. The CGF threshold $R_\cgf(\lambdagro)=e^{1/\rho_{\lambdagro}}$ and the drift threshold~\eqref{eq:drift-reward-threshold} use $\lambdagro$, and the plug-in uses the threshold rule~\eqref{eq:plugin-threshold}. Ratios are estimated from $20{,}000$ simulations with a cap of $2500$ rounds. }
\label{fig:efficiency}
\end{figure*}

We use Gaussian observations $X_i\sim N(\theta,0.5^2)$ with the exponential reward~\eqref{eq:example-process}, and Bernoulli observations with the betting reward above. The growth-optimal parameters are $\lambdagro=4\theta$ and $\lambdagro=(p-m)/(m(1-m))$, respectively. For Bernoulli data, Figure~\ref{fig:efficiency} also includes a Hoeffding baseline, which uses the sub-Gaussian reward $C\prod_{i\leq t}\exp\{\lambda(X_i-m)-\lambda^2/8\}$, valid for any data in $[0,1]$ with mean at most $m$. Its drift $\lambda(p-m)-\lambda^2/8$ depends only on the mean, so, as with the Gaussian reward, an agent who knows only $p$ can implement the growth-optimal rule exactly, with $\lambda=4(p-m)$ and the drift threshold~\eqref{eq:drift-reward-threshold}. The price is a smaller drift, and hence lower utility. Appendix~\ref{app:reward-families} quantifies the gap using Pinsker's inequality, derives the CGF-optimal parameters, and specifies the plug-in rules.

In Figure~\ref{fig:efficiency}, the CGF-optimal policy~\eqref{eq:cgf-policy} is within $2\%$ of the  Bellman benchmark throughout the Gaussian settings and within $1.2\%$ throughout the Bernoulli settings. The drift rule~\eqref{eq:drift-reward-threshold} performs similarly for patient agents, but loses more with stronger discounting. The plug-in rule~\eqref{eq:plugin-threshold} can lose substantially for weak signals or strong discounting.

For the bounded Bernoulli increments in Figure~\ref{fig:efficiency}, Proposition~\ref{prop:cgf-guarantee} quantifies the loss from overshoot. Gaussian log-increments are unbounded above, but equation~\eqref{eq:deferred-overshoot-moment} in Appendix~\ref{app:proofs} bounds the loss using the second moment of the overshoot. As hinted at by these results, the performance of our strategies can decline substantially when a single observation can carry the reward far past the threshold.

To give a concrete example, consider Bernoulli data with rare successes ($m=0.01$, $p=0.02$, $\delta=0.9$). Under $\lambda_\cgf\approx54.5$, one success multiplies the evidence by about $55$, a log-jump of $4.0$ against the log-evidence threshold $1/\rho^*\approx1.17$ in~\eqref{eq:cgf-policy}, so $q\approx0.14$. The CGF-optimal rule earns $0.184$, whereas the smaller bet $\lambda=15.5$ with its exact threshold $\beta^\star(15.5)$ from Proposition~\ref{prop:threshold} earns $0.261$. So the CGF-optimal rule loses at least $29.4\%$ of the optimal value. More details are provided in Appendix~\ref{app:large-jump}.

    \section{Conclusion}

    We have extended the framework of principal-agent hypothesis testing to settings with repeated and adaptive experimentation. We introduced sequential statistical contracts and showed how a principal can use e-processes to design contracts which deter the entry of null agents. We then focused on the agent, noting that techniques from stochastic control theory can be leveraged to optimize the reward process once they've opted into the contract. Such techniques are not closed-form, however, and require that the agent possess full knowledge of their distribution to implement them. We therefore provide several approximations which are more transparent and can be implemented with less information.

    \paragraph{Acknowledgements.}
    BC and AR acknowledge support from NSF grants IIS-2229881 and DMS-2310718.

	{
    \ifarxiv
    \footnotesize
		\bibliographystyle{plainnat}
        \fi
		\bibliography{main}
	}

    \ifarxiv\else
    \clearpage
    \fi

	\appendix

	\section{Omitted Proofs}
\label{app:proofs}

We give the proofs of the four propositions in the main text. When the parameter $\lambda$ is fixed, we suppress its subscript in the increments, partial sums, and so on.

\begin{proof}[Proof of Proposition~\ref{prop:null-deterrence}]
        If $\calC$ is ND, then by definition $\E_\theta[R_{\tau\wedge \sigma}] \leq C$ for any stopping time $\tau\geq1$. The bound also holds when $\tau$ may be zero: since $R_0=0$ and rewards are nonnegative, $R_{\tau\wedge\sigma}\leq R_{(\tau\vee1)\wedge\sigma}$. Dividing by $C>0$ shows that $(R_{t\wedge\sigma}/C)$ is an e-process for $\Theta_0$.

        Conversely, on the deferred schedule, if $(R_{t\wedge\sigma}/C)$ is an e-process, then $\E_\theta[R_{\tau\wedge\sigma}]\leq C$ for all stopping times $\tau$, which is precisely ND.

        On the incremental schedule, suppose every $(R_t)\in\calR$ is nondecreasing, so that $R_\infty^\sigma=\lim_t R_{t\wedge\sigma}$ exists in $[0,\infty]$ for each reward process. If $\calC$ is ND, then taking the deterministic stopping times $\tau=n$ gives $\E_\theta[R_{n\wedge\sigma}]\leq C$, and monotone convergence yields $\E_\theta[R_\infty^\sigma]\leq C$. Conversely, if $\E_\theta[R_\infty^\sigma]\leq C$, then $R_{\tau\wedge\sigma}\leq R_\infty^\sigma$ for every stopping time $\tau$, so $\E_\theta[R_{\tau\wedge\sigma}]\leq C$.

        In particular, if $(R_t/C)$ is an e-process for every reward process, then $\E_\theta[R_{n\wedge\sigma}]\leq C$ for every $n$, since $n\wedge\sigma$ is a stopping time. Monotone convergence gives the required limit bound, hence ND. It remains to verify the adjuster construction.
        Let $(E_t)$ be an e-process for
        $\Theta_0$. By an equivalent definition of an e-process~\citep{ramdas2023game}, for each
        $\theta \in \Theta_0$ there exists a test supermartingale $(M_t^\theta)_{t \ge 0}$ under $P_\theta$
        such that $E_t \le M_t^\theta$ $P_\theta$-a.s.\ for all $t\geq1$.
        Moreover, since $A$ is an adjuster, there exists a test supermartingale $(Y_t^\theta)$ under $P_\theta$ such that $A(\max_{0\leq k\leq t} M_k^\theta) \leq Y_t^\theta$ $P_\theta$-a.s.
        Therefore, since $A$ is nondecreasing (and since $E_t^*=1\vee\max_{k\leq t}E_k\geq 1$ lies in the domain of $A$ and $M_0^\theta=1$), for any $t\geq 1$,
        \begin{equation*}
            A(E_t^*) \leq A\left(\max_{0\leq k\leq t} M_k^\theta\right) \leq Y_t^\theta \quad P_\theta\text{-a.s.},
        \end{equation*}
        hence $\E_\theta[A(E_\tau^*)]\leq \E_\theta[Y_\tau^\theta]\leq 1$. We conclude that $(A(E_t^*))_{t\geq 1}$ is an e-process for $\Theta_0$, immediately implying that $(CA(E_t^*))_{t\geq 1}$ is null-deterring.
    \end{proof}

{
The following identity supplies the overshoot correction used in the proofs below and extends the performance guarantees to unbounded increments.

\begin{lemma}
\label{lem:deltaTk-approx}
Fix $\lambda$ with $\psi_\lambda(\rho_\lambda)=\log(1/\delta)$ for some $\rho_\lambda>0$. Let $P_{\lambda,\rho_\lambda}$ be the tilted law defined by
\[
\frac{\d P_{\lambda,\rho_\lambda}}{\d P_\theta}\bigg|_{\calF_n}
=\exp\{\rho_\lambda W_n(\lambda)-n\log(1/\delta)\}.
\]
For $\kappa\geq0$, the crossing time $T_\lambda(\kappa)$ is finite under this law. Write $O_\lambda(\kappa)=W_{T_\lambda(\kappa)}(\lambda)-\kappa$ for the overshoot. Then
\begin{equation}
\label{eq:deltaTk-identity}
H_\lambda(\kappa)=e^{-\rho_\lambda\kappa}\E_{P_{\lambda,\rho_\lambda}}
[e^{-\rho_\lambda O_\lambda(\kappa)}].
\end{equation}
More generally, for any nonnegative measurable $g$,
\begin{equation}
\label{eq:tilted-crossing-payoff}
\begin{aligned}
&\E_\theta[\delta^{T_\lambda(\kappa)}g(W_{T_\lambda(\kappa)}(\lambda))
 \ind{T_\lambda(\kappa)<\infty}]\aaaibreak[\qquad]=e^{-\rho_\lambda\kappa}\E_{P_{\lambda,\rho_\lambda}}
 [e^{-\rho_\lambda O_\lambda(\kappa)}g(\kappa+O_\lambda(\kappa))].
\end{aligned}
\end{equation}
The formula extends to signed $g$ whenever one of its weighted positive or negative parts is integrable.
\end{lemma}

\begin{proof}
Suppress $\lambda$ and write $\rho=\rho_\lambda$, $T=T_\lambda(\kappa)$, and $M_n=\delta^ne^{\rho W_n}$. Tilting the one-observation law by $\delta e^{\rho\Delta_1}$ and taking its iid product defines $P_\rho$, since $\delta\E_\theta[e^{\rho\Delta_1}]=1$. Its density on $\calF_n$ is $M_n$. Hence $\E_{P_\rho}[e^{-\rho W_n}]=\delta^n$, and
\[
P_\rho(T>n)\leq P_\rho(W_n<\kappa)
\leq e^{\rho\kappa}\delta^n\longrightarrow0.
\]
Changing measure on each event $\{T=n\}$ gives
\[
\E_\theta[\delta^Tg(W_T)\ind{T=n}]
=\E_{P_\rho}[e^{-\rho W_T}g(W_T)\ind{T=n}].
\]
Sum over $n\geq0$ and use $W_T=\kappa+O$ to obtain~\eqref{eq:tilted-crossing-payoff}. Taking $g\equiv1$ gives~\eqref{eq:deltaTk-identity}. Applying the formula separately to positive and negative parts gives the signed extension.
\end{proof}

For $\kappa>0$ and $g(w)=\log C+w$, both weighted parts are integrable on $w\geq\kappa$, because $(|\log C|+w)e^{-\rho_\lambda w}$ is bounded. Thus
\begin{equation}
\label{eq:deferred-overshoot-identity}
J_\lambda(\kappa)=e^{-\rho_\lambda\kappa}
\E_{P_{\lambda,\rho_\lambda}}\!\left[
e^{-\rho_\lambda O_\lambda(\kappa)}(\log C+\kappa+O_\lambda(\kappa))\right].
\end{equation}
In particular, if $\psi_{\lambda_\cgf}(\rho^*)=\log(1/\delta)$ and $\log C<1/\rho^*$, setting $\kappa=1/\rho^*-\log C$ and $O=O_{\lambda_\cgf}(\kappa)$ gives
\begin{equation}
\label{eq:deferred-moment-ratio}
\frac{J_{\lambda_\cgf}(\kappa)}{U_{\dfrd}(C)}
=\E_{P_{\lambda_\cgf,\rho^*}}[(1+\rho^*O)e^{-\rho^*O}].
\end{equation}
This identity also applies to unbounded increments.

Under the same assumptions, since $0\leq1-(1+x)e^{-x}=\int_0^x te^{-t}\d t\leq x^2/2$ for $x\geq0$, it gives
\begin{equation}
\label{eq:deferred-overshoot-moment}
\begin{aligned}
&0\leq1-\frac{J_{\lambda_\cgf}(\kappa)}{U_{\dfrd}(C)}
\aaaibreak\leq\frac{(\rho^*)^2}{2}\E_{P_{\lambda_\cgf,\rho^*}}[O^2].
\end{aligned}
\end{equation}
Since $J_{\lambda_\cgf}(\kappa)\leq V_\theta^\star(C)\leq U_{\dfrd}(C)$, the same bound controls the relative loss against the optimal adaptive strategy. Near-optimality thus follows when $(\rho^*)^2\E_{P_{\lambda_\cgf,\rho^*}}[O^2]$ is small, without a bounded-increment assumption.
For Gaussian increments, this tilted second moment is finite. Indeed, tilting preserves normality, the lemma's geometric tail bound gives a finite tilted mean crossing time, and $O^2\leq\sum_{i=1}^{T_{\lambda_\cgf}(\kappa)}(\Delta_i(\lambda_\cgf)^+)^2$ has finite tilted expectation by independence.

\begin{proof}[Proof of Proposition~\ref{prop:crossing-comparison}]
For a predictable strategy $\pi$, write $M_t=\delta^t(E_t^\pi)^s$. Independence of $X_t$ from $\calF_{t-1}$ and the conditional moment bound give
\[
\E_\theta[M_t\mid\calF_{t-1}]
=M_{t-1}\delta\E_\theta[f_{\lambda_t}(X_t)^s\mid\calF_{t-1}]
\leq M_{t-1}.
\]
Thus $(M_t)$ is a nonnegative supermartingale with initial value one, and obeys $\E_\theta[M_T]\leq 1$ for $T = T_\pi(\kappa)$. If $T$ is finite, $E_T^\pi\geq e^\kappa$ by definition, hence $H_\pi(\kappa)\leq e^{-s\kappa}$.

Under the additional assumptions for $\lambda_0$, Lemma~\ref{lem:deltaTk-approx} applies with $\rho_{\lambda_0}=s$. Its overshoot satisfies $0\leq O_{\lambda_0}(\kappa)\leq h$, so
\[
H_{\lambda_0}(\kappa)
=e^{-s\kappa}\E_{P_{\lambda_0,s}}[e^{-sO_{\lambda_0}(\kappa)}]
\geq e^{-s(\kappa+h)}.
\]
The remaining inequality follows because a fixed parameter is a predictable strategy.
\end{proof}

\begin{proof}[Proof of Proposition~\ref{prop:incremental-guarantee}]
The baseline $CA(1)$ is paid in round one, and each increase $C\,\d A(a)$ is paid at its first evidence crossing. Tonelli's theorem gives~\eqref{eq:incremental-adjuster-value}, since these payments are nonnegative. Integrating the upper crossing bound gives $J_{\mathrm{inc}}(\pi)\leq U_A$ for every predictable policy.

Under the lower-bound assumptions, Proposition~\ref{prop:crossing-comparison} gives
\[
\begin{aligned}
J_{\mathrm{inc}}(\lambda_\cgf)
&\geq\delta CA(1)+e^{-\rho^*h}C\int_{(1,\infty)}a^{-\rho^*}\,\d A(a)\aaaibreak\geq e^{-\rho^*h}U_A.
\end{aligned}
\]
The middle inequality in~\eqref{eq:incremental-guarantee} follows because the fixed CGF policy is admissible.
\end{proof}

\begin{proof}[Proof of Proposition~\ref{prop:cgf-guarantee}]
Write $c=\log C$ and $Y_t=c+W_t$, so that $Y_t=\log R_t$ for $t\geq1$. For an adaptive strategy, abbreviate $\Delta_t(\lambda_t)$ as $\Delta_t$. Let $L_t=\delta^te^{\rho^* W_t}$. Fix any strategy. Since $\lambda_t$ is $\calF_{t-1}$-measurable, $X_t$ is independent of $\calF_{t-1}$, and $\psi_\lambda(\rho^*)\leq\log(1/\delta)$ for every $\lambda$ (Section~\ref{sec:discounted-evidence}),
\begin{equation*}
    \E_\theta[\delta e^{\rho^*\Delta_t}\mid\calF_{t-1}]=\delta e^{\psi_{\lambda_t}(\rho^*)}\leq1,
\end{equation*}
so $(L_t)$ is a nonnegative supermartingale with $L_0=1$. Thus $\E_\theta[L_\tau]\leq 1$ for any stopping time $\tau\geq 1$. Since $ze^{-\rho^* z}\leq1/(e\rho^*)$ for all $z$,
\begin{equation*}
    G_\tau^\dfrd
    =C^{\rho^*}L_\tau Y_\tau e^{-\rho^*Y_\tau}\ind{\tau<\infty}
    \leq \frac{C^{\rho^*}L_\tau\ind{\tau<\infty}}{e\rho^*}.
\end{equation*}
The positive part of the utility is therefore integrable. If its negative part is not integrable, the upper bound is immediate. Otherwise, set $d=\E_\theta[\delta^\tau\ind{\tau<\infty}]\in[0,\delta]$. If $d=0$, the utility is zero. For $d>0$, Jensen's inequality under the probability measure with density $\delta^\tau\ind{\tau<\infty}/d$ gives
\begin{align*}
    \E_\theta[G_\tau^\dfrd]
    &\leq\frac{d}{\rho^*}\log\left(\frac{\E_\theta[\delta^\tau R_\tau^{\rho^*}\ind{\tau<\infty}]}{d}\right)\aaaibreak\leq dc-\frac{d\log d}{\rho^*},
\end{align*}
where the second inequality uses $\E_\theta[\delta^\tau R_\tau^{\rho^*}\ind{\tau<\infty}]=C^{\rho^*}\E_\theta[L_\tau\ind{\tau<\infty}]\leq C^{\rho^*}$. Enlarging the range to $0<d\leq1$ and maximizing gives $d=\min(1,C^{\rho^*}/e)$ and the bound $U_{\dfrd}(C)$, which is positive and also covers $d=0$.

Under the additional assumptions for the lower bound, set $\kappa=1/\rho^*-c$ and $O=O_{\lambda_\cgf}(\kappa)$. The overshoot identity~\eqref{eq:deferred-moment-ratio} gives
\[
\begin{aligned}
\frac{J_{\lambda_\cgf}(\kappa)}{U_{\dfrd}(C)}
&=\E_{P_{\lambda_\cgf,\rho^*}}[(1+\rho^*O)e^{-\rho^*O}]\aaaibreak\geq(1+\rho^*h)e^{-\rho^*h}=q,
\end{aligned}
\]
because $0\leq O\leq h$ under the tilted law and $(1+x)e^{-x}$ is nonincreasing for $x\geq0$. The threshold rule is admissible, so $J_{\lambda_\cgf}(\kappa)\leq V_\theta^\star(C)$.
\end{proof}
}

\section{Additional Results}
\label{app:additional-results}

{
\subsection{Bellman and Snell characterizations}
\label{sec:full-knowledge}

Once the agent has chosen a reward process, they face a joint optimization over parameter selection and stopping, which can be framed as a stochastic control problem. Let $v(r,t)$ denote the agent's utility from collecting reward $r$ at time $t$. For deferred rewards with $R_t>0$ for $t\geq1$, let $W_t=\log(R_t/C)$ and define, after round $t\geq1$ while the contract is active,
\begin{equation}
\label{eq:bellman-value}
F(w,t)=\sup_{\pi,\tau\geq t}\E_{\theta,\pi}
[v(R_{\tau\wedge\sigma},\tau\wedge\sigma)\mid W_t=w].
\end{equation}
Here $\pi$ ranges over predictable parameter policies and $\tau$ over almost surely finite stopping times. The reward and time arguments are frozen when the principal stops at $\sigma$.

The Bellman equation characterizes $F$ under a Markov assumption: the conditional law of $W_{t+1}$ given $\calF_t$ and $\lambda_{t+1}$ depends only on $(W_t,t,\lambda_{t+1})$, and the principal stops at the first entry of $(W_t,t)$ into some termination set $D$, that is, $\sigma=\inf\{t:(W_t,t)\in D\}$. Examples include a deterministic horizon $\sigma=N$, a rule that stops once the reward reaches a target or leaves an interval, a time-varying boundary $\sigma=\inf\{t:W_t\geq b(t)\}$, and a growth-rate rule $\sigma=\inf\{t:W_t/t\leq\gamma\}$. A path-dependent rule, or one driven by the principal's private information (such as a budget), is not Markov in $(W_t,t)$. Augmenting the state can restore the Markov property, but at the cost of a larger state, or of state variables the agent cannot observe.

\begin{proposition}[Bellman characterization]
\label{prop:bellman}
Suppose the Markov assumption holds and the stopped payoffs are uniformly integrable over admissible policies and stopping times. For $t\geq1$ on $\{t<\sigma\}$,
\begin{equation}
\label{eq:bellman}
\begin{aligned}
&F(w,t)=\max\big\{v(Ce^w,t),\ \aaaibreak[\quad]\sup_{\lambda\in\Lambda}\E_\theta[F(W_{t+1},t+1)\mid W_t=w,\lambda_{t+1}=\lambda]\big\},
\end{aligned}
\end{equation}
with $F(w,t)=v(Ce^w,t)$ at termination. The value at entry is $\sup_{\lambda_1\in\Lambda}\E_\theta[F(W_1,1)]$, since the agent has already opted in. If the supremum admits a measurable maximizing selector and the resulting first stopping time is almost surely finite, selecting that maximizer while the continuation value exceeds the stopping payoff is optimal.
\end{proposition}

This follows from the principle of optimality and optional stopping under uniform integrability~\citep[Chapter~3]{peskir2006optimal}. The selector and finiteness conditions are needed only for an optimal strategy to exist, not for the Bellman equation itself. The numerical benchmarks use bounded action sets with integrable envelopes for the log-increments, which ensure uniform integrability of discounted log payoffs. Incremental rewards require a state that also records past payments, since discounted receipts depend on when each payment was made; for milestone contracts, current evidence and the next unpaid milestone suffice (Appendix~\ref{app:incremental-computation}).

The Bellman equation is rarely solvable in closed form. Even for Gaussian data it is a nonlinear integral equation, coupled with a state-dependent optimization over $\lambda$, and must be solved numerically. In practice, however, e-processes are seldom optimized round by round. One either commits in advance to a predictable plug-in rule $\lambda_t=g_t(X_1,\dots,X_{t-1})$~\citep{waudby2024estimating} or mixes over $\lambda$ with a data-free distribution. Either way, the reward process is determined by the data, and only the stopping decision remains. For $t\geq1$, write $Y_t=v(R_{t\wedge\sigma},t\wedge\sigma)$ and define its Snell envelope, the smallest supermartingale that dominates $(Y_t)$, by
\begin{equation}
\label{eq:snell-envelope}
S_t=\esssup_{\tau\geq t,\ \tau<\infty\ \mathrm{a.s.}}
\E_\theta[Y_\tau\mid\calF_t].
\end{equation}

\begin{theorem}[Snell characterization]
\label{thm:snell}
Suppose $\sigma\geq1$ is an $(\calF_t)$-stopping time and $\E_\theta[\sup_{t\geq1}|Y_t|]<\infty$. If $\tau^*=\inf\{t\geq1:Y_t=S_t\}\wedge\sigma$ is almost surely finite, it attains $\sup_{1\leq\tau<\infty\ \mathrm{a.s.}}\E_\theta[Y_\tau]$.
\end{theorem}

\begin{proof}
For $t\geq1$, the recursion $S_t=\max\{Y_t,\E_\theta[S_{t+1}\mid\calF_t]\}$ makes $(S_{t\wedge\tau^*})_{t\geq1}$ a uniformly integrable martingale. Thus $\E_\theta[Y_{\tau^*}]=\E_\theta[S_1]$, while domination and optional stopping give $\E_\theta[Y_\tau]\leq\E_\theta[S_1]$ for every admissible $\tau\geq1$.
\end{proof}

The integrability condition can be weakened to uniform integrability of $\{Y_\tau:\tau<\infty\text{ a.s.}\}$, the usual class-D assumption~\citep[Chapter~1]{peskir2006optimal}. For a fixed parameter with $\E_\theta|\Delta_1(\lambda)|<\infty$, it holds for discounted log utility, since $\sup_{t\geq1}|Y_t|\leq|\log C|+\sum_{i\geq1}\delta^i|\Delta_i(\lambda)|$, whose expectation is at most $|\log C|+\delta\E_\theta|\Delta_1(\lambda)|/(1-\delta)$. For conditions under which the expected log utility of an e-value is well-defined, see \Citet[Lemma~2.5]{deroos2025rao}.

Unlike Proposition~\ref{prop:bellman}, Theorem~\ref{thm:snell} does not require $\sigma$ to be Markov, only an $(\calF_t)$-stopping time. For instance, a principal who stops once the reward falls more than a fraction $\eps$ below its running maximum, $\sigma=\inf\{t\geq1:R_t\leq(1-\eps)\max_{s\leq t}R_s\}$, uses a path-dependent rule that cannot be written in terms of $(W_t,t)$. The Snell envelope thus trades per-round optimization of $\lambda_t$ for more general principal stopping rules. Both characterizations generally require computing a value function, and both also apply to other adapted payoffs, including cumulative discounted incremental receipts.

Section~\ref{sec:cgf-schedules} also allows $\tau=\infty$, with payoff zero. For discounted log utility under the integrability conditions above, this changes neither value. Indeed, $\sup_t|Y_t|$ is then integrable and, by the strong law of large numbers, $Y_t\to0$ almost surely on $\{\sigma=\infty\}$, so dominated convergence gives $\E_\theta[Y_{\tau\wedge n}]\to\E_\theta[Y_\tau]$ for every stopping time $\tau$.

\subsection{Exact fixed-parameter stopping value}
\label{app:fixed-stopping-details}
For a fixed parameter, the optimal stopping boundary has an exact series representation~\citep{darling1972optimal,novikov2005effective}.

\begin{proposition}
\label{prop:threshold}
Consider deferred rewards $R_0=0$ and $R_t=CE_t$ for $t\geq1$, with $\sigma=\infty$ and utility $G_t^\dfrd=\delta^t\log R_t$, with $G_\infty^\dfrd=0$. Fix $\lambda$ with $\E_\theta|\Delta_1(\lambda)|<\infty$. Set $S_n(\lambda)=\sum_{i=1}^n\Delta_i(\lambda)$ and
\[
\begin{gathered}
\beta^\star(\lambda)=\sum_{n=1}^\infty\frac{\delta^n}{n}\E_\theta[(S_n(\lambda))^+],
\ifarxiv\qquad\else\\\fi
\tau_\beta(\lambda)=\inf\{t\geq1:\log R_t(\lambda)\geq\beta^\star(\lambda)\}.
\end{gathered}
\]
The rule $\tau_\beta(\lambda)$ attains the supremum of expected utility over all stopping times $\tau\geq1$ at this parameter, without a positive-drift assumption.
\end{proposition}

Under the assumptions of Proposition~\ref{prop:threshold}, let $N_\delta$ be independent of the observations, with $\Pr(N_\delta=n)=(1-\delta)\delta^n$ for $n=0,1,\ldots$. Since $\Pr(N_\delta\geq t)=\delta^t$, discounting is equivalent to crossing before this deadline:
\begin{equation}
\label{eq:crossing-random-deadline}
H_\pi(\kappa)=\Pr_{\theta,\pi}\!\left(\max_{0\leq t\leq N_\delta}W_t^\pi\geq\kappa\right).
\end{equation}
For a fixed parameter set $M_\lambda=\max_{0\leq t\leq N_\delta}S_t(\lambda)$. The boundary is the expected record high, $\beta^\star(\lambda)=\E_\theta[M_\lambda]$. Since the agent must run the first trial, the exact value is
\begin{equation}
\label{eq:exact-fixed-value}
\begin{aligned}
V_\lambda(C)&:=\sup_{\tau\geq1}\E_\theta[G_\tau^\dfrd]\aaaibreak=\delta\E_\theta[(\log C+\Delta_1(\lambda)+M_\lambda'-\beta^\star(\lambda))^+],
\end{aligned}
\end{equation}
where $M_\lambda'$ is an independent copy of $M_\lambda$, independent of $\Delta_1(\lambda)$. When $\log C\leq\beta^\star(\lambda)$, this reduces to $\E_\theta[(\log C+M_\lambda-\beta^\star(\lambda))^+]$; this includes the experiments with $C=1$.

\begin{proof}[Proof of Proposition~\ref{prop:threshold}]
    Write $x=\log C$, so the payoff at time $t\geq1$ is $\delta^t(x+S_t)$. Use the independent geometric deadline $N_\delta$ from~\eqref{eq:crossing-random-deadline} and let $M=\max_{0\leq k\leq N_\delta}S_k$. Since $M\leq\sum_{k\leq N_\delta}\Delta_k^+$, we have $\E[M]\leq\E[N_\delta]\E[\Delta_1^+]=\frac{\delta}{1-\delta}\E[\Delta_1^+]<\infty$.

    Spitzer's identity~\citep{spitzer1956combinatorial} gives $\E[\max_{0\leq k\leq n}S_k]=\sum_{k=1}^n\E[S_k^+]/k$. Averaging over $N_\delta$ and exchanging the order of summation,
    \begin{align*}
        \E[M]&=\sum_{n\geq0}(1-\delta)\delta^n\sum_{k=1}^n\frac{\E[S_k^+]}{k}\aaaibreak=\sum_{k\geq1}\frac{\E[S_k^+]}{k}\sum_{n\geq k}(1-\delta)\delta^n\aaaibreak=\sum_{k\geq1}\frac{\delta^k}{k}\E[S_k^+]=\beta^\star.
    \end{align*}
    Define the candidate value function
    \begin{equation*}
        U(y):=\E\big[(y+M-\beta^\star)^+\big],\qquad y\in\Re.
    \end{equation*}

    \emph{Step 1: $U$ dominates the payoff, with equality above $\beta^\star$.} By Jensen's inequality, $U(y)\geq (y+\E M-\beta^\star)^+\geq y$. If $y\geq\beta^\star$ then $y+M-\beta^\star\geq 0$ (as $M\geq 0$), so $U(y)=y+\E M-\beta^\star=y$.

    \emph{Step 2: a one-step recursion.} By the memoryless property of the geometric law, with probability $1-\delta$ we have $N_\delta=0$ and $M=0$, while conditional on $N_\delta\geq 1$ we have $M\overset{d}{=}\max(0,\Delta_1+M')$, where $M'\overset{d}{=}M$ is independent of $\Delta_1$. Hence
    \begin{align*}
        U(y)&=(1-\delta)(y-\beta^\star)^+\aaaibreak[\quad]+\delta\,\E\big[(y-\beta^\star+\max(0,\Delta_1+M'))^+\big].
    \end{align*}
    If $y\leq\beta^\star$, then $(y-\beta^\star+\max(0,z))^+=(y-\beta^\star+z)^+$ for every $z$ (both vanish when $z\leq 0$), so $U(y)=\delta\E[U(y+\Delta_1)]$. If $y\geq\beta^\star$, then $(y-\beta^\star+\max(0,z))^+\geq(y-\beta^\star+z)^+$, so $U(y)\geq\delta\E[U(y+\Delta_1)]$. Thus $(\delta^tU(x+S_t))_{t\geq1}$ is a nonnegative supermartingale, and a martingale from time one up to the hitting time $T:=\inf\{t\geq1:x+S_t\geq\beta^\star\}$.

    \emph{Step 3: upper bound.} For any stopping time $\tau\geq1$, Step 1 and optional stopping from time one (via Fatou's lemma) give
    \[
    \E[G_\tau^\dfrd]\leq\E[\delta^\tau U(x+S_\tau)\ind{\tau<\infty}]
    \leq\delta\E[U(x+\Delta_1)],
    \]
    where the stopped term is defined as zero on $\{\tau=\infty\}$.

    \emph{Step 4: attainment.} The stopped process from Step 2 is an integrable martingale from time one at every finite horizon. On $\{T\leq n\}$, Step 1 gives $U(x+S_T)=x+S_T$. Consequently, for $n\geq1$,
    \begin{align*}
    \delta\E[U(x+\Delta_1)]
    &=\E[\delta^T(x+S_T)\ind{T\leq n}]\aaaibreak[\quad]+\delta^n\E[U(x+S_n)\ind{T>n}].
    \end{align*}
    On $\{T>n\}$, $x+S_n<\beta^\star$, so monotonicity of $U$ bounds the second term by $\delta^nU(\beta^\star)=\delta^n\beta^\star\to0$. The first term converges to $\E[G_T^\dfrd]$ by monotone convergence, since $x+S_T\geq\beta^\star\geq0$ at every finite crossing. Thus $T$ attains the upper bound $\delta\E[U(x+\Delta_1)]$, proving~\eqref{eq:exact-fixed-value}. If $x\leq\beta^\star$, Step 2 also gives $\delta\E[U(x+\Delta_1)]=U(x)$.
    \end{proof}

\subsection{A limitation of the adjuster construction}
\label{app:adjuster-counterexample}

Here we observe that not all nondecreasing e-processes come from adjusters.

\begin{proposition}
\label{prop:adjuster-nonrepresentation}
There exists a bounded nondecreasing e-process $(R_t)_{t\geq0}$ with $R_0=0$ for a singleton null $\{P\}$ that cannot be represented as
\[
R_t=A(E_t^*),\qquad E_t^*=1\vee\max_{1\leq s\leq t}E_s,\quad t\geq1,
\]
for any deterministic adjuster $A$ and any e-process $(E_t)$ for $P$ on the same filtration. In fact, no such pair satisfies $R_t\leq A(E_t^*)$ almost surely for every $t\geq1$.
\end{proposition}

\begin{proof}
Let $X$ be uniform on $\{0,1,2\}$ under $P$, with $\calF_0$ trivial and $\calF_t=\sigma(X)$ for $t\geq1$. Set $R_0=0$ and $R_t=X$ for $t\geq1$. This process is bounded and nondecreasing. For every stopping time $\tau$, $0\leq R_\tau\leq X$, so $\E_P[R_\tau]\leq\E_P[X]=1$. Thus $(R_t)$ is an e-process and defines a null-deterring incremental contract with entry cost $C=1$.

Suppose $X=A(1\vee E_1)$ almost surely. Since $P(X=0)>0$, nonnegativity and monotonicity of $A$ force $A(1)=0$. Moreover, every adjuster satisfies $A(y)\leq y$ for $y\geq1$, because
\[
1\geq\int_1^\infty\frac{A(u)}{u^2}\,\d u
\geq\int_y^\infty\frac{A(y)}{u^2}\,\d u
=\frac{A(y)}y.
\]
On $\{X=1\}$, we must have $E_1>1$: otherwise $A(1\vee E_1)=A(1)=0$. On $\{X=2\}$, we have $E_1>1$ and $2=A(E_1)\leq E_1$. On $\{X=0\}$, nonnegativity gives $E_1\geq0$. Hence $E_1\geq X$ almost surely, with strict inequality on an event of probability $1/3$. It follows that $\E_P[E_1]>\E_P[X]=1$, contradicting the e-process property.

Finally, if $X\leq A(1\vee E_1)$ almost surely, the adjuster guarantee in Proposition~\ref{prop:null-deterrence} gives $1=\E_P[X]\leq\E_P[A(1\vee E_1)]\leq1$. Equality of expectations and the assumed domination force $X=A(1\vee E_1)$ almost surely, which was just ruled out.
\end{proof}

}

\section{Comparison to the Sequential Contract of Bates et al.}
\label{app:bates-comparison}

As we discussed in the introduction, \citet{bates2022principal} also provided a sequential version of their one-shot contract. We state it here:

    \begin{tcolorbox}[
		colback=blue!8!white,
		colframe=white,
		boxrule=0pt,
		sharp corners,
		left=6pt,
		right=6pt,
		top=6pt,
		bottom=6pt,
		breakable=true
		]
		\textbf{\citet{bates2022principal} multi-round contract}
		\begin{enumerate}[nosep,topsep=0.2cm]
        \item The initial license value is $L(0) = 0$.
        \item For $t=1,2,\dots,T$:
        \begin{enumerate}
            \item The agent chooses to withdraw $P (t) \in (0, L(t-1))$ from their license, or continue without a withdrawal.
            \item The agent may pay $C_t$ to run stage $t$ of the trial, in which case:
            \begin{enumerate}
                \item The agent selects an update function $f_t$ from the menu $\mathcal{U}_t$
                \item The trial produces evidence $Z_t \sim P_\theta$
                \item The license value is updated as $L(t) = (L(t-1) + C_t - P (t)) \cdot f_t (Z_t)$.
            \end{enumerate}
            \item Otherwise, license value is updated as $L(t) = L(t -1) - P (t)$.
        \end{enumerate}
		\end{enumerate}
	\end{tcolorbox}

    While their motivation is similar to ours, the two contracts differ  in their treatment of costs, payments, horizons, and the agent's optimization problem. Let us draw attention to each of these features in turn.

\textit{Per-round costs and payments.}
    First, the contract above uses per-round costs, and allows the agent to withdraw money from the license won so far. This can be beneficial for some applications, as it makes the cost structure explicit and part of the agent's strategy. In particular, the contract is well-suited to settings in which continued experimentation
requires a sequence of distinct investments and the agent actively manages both investment and consumption.

We instead model $C$ as the up-front cost of mounting the research program. Any costs of delay or continuation are represented through the
agent's utility $v(R_t,t)$ rather than through additional payments in the
contract. In other words, per-round costs are possible in our setting, though they are modeled only on the agent's side---they are not part of the structure of the contract itself.

\textit{Finite-horizon and stopping times.}
Per-round costs make an unbounded horizon more delicate. The agent's net-profit supermartingale need not be nonnegative, so optional stopping does not apply to arbitrary stopping times without additional admissibility conditions. (For instance, if cumulative investment is bounded by $B$, the net profit is bounded below by $-B$, and optional stopping can be applied to the shifted process.) Consequently, Bates et al.\ work with a finite number $T$ of possible rounds. This fixed horizon allows the problem to be solved by backwards induction.
Our framework, meanwhile, allows an unbounded horizon and introduces stopping rules for both the principal and the agent.

\textit{The agent's best response.} Both we and Bates et al.\ study the agent's best response, and both do so via dynamic programming. Bates et al. formulate the finite-horizon contract as a Markov decision
process whose state includes the current time and license value. At each stage, the agent chooses whether to stop and, if continuing, which e-value update to use. Our Bellman formulation is analogous, but permits an unbounded horizon, an explicit principal stopping rule, and general time-dependent utility. For a fixed reward process, we additionally characterize the optimal
stopping time through its Snell envelope.

Overall, the two frameworks should be viewed as complementary extensions of the single-shot contract. Bates et al. emphasize a finite sequence of costly investments with flexible withdrawals and reinvestment. We emphasize a potentially indefinite stream of evidence,
a fixed entry cost, principal and agent stopping, and approximations to the optimal stopping rule under different levels of agent information.

\ifarxiv\else

\fi

    \section{Experimental Details}
\label{app:sims}
\label{app:reward-families}
\label{app:incremental-computation}
\label{app:deferred-simulations}
\label{app:large-jump}

Code to replicate the experiments is available at \url{https://github.com/bchugg/statcontracts}.

For Figure~\ref{fig:incremental-milestones}, we use Bernoulli observations and the milestone allocations in Section~\ref{sec:incremental-experiment}. For the betting multiplier $f_\lambda(X)=1+\lambda(X-m)$, the log-increment CGF is
\[
\psi_\lambda(s)=\log\big[p(1+\lambda(1-m))^s+(1-p)(1-\lambda m)^s\big].
\]
We compute the CGF-optimal parameter by numerically solving $\sup_{\lambda\in\Lambda}\psi_\lambda(\rho^*)=\log(1/\delta)$; a prescribed fixed bet uses its own root $\rho_\lambda$. All reported Bernoulli settings satisfy $\delta p>m$, so $\rho^*\in(0,1)$ and an interior CGF-optimal bet exists. The mean plug-in starts at $\lambda_1=0.1$, then uses $\widetilde p_t=\min\{\bar X_t,0.99\}$ and $\lambda_{t+1}=(\widetilde p_t-m)^+/(m(1-m))$, continuing to sample when the bet is zero.

A policy's value is $J_{\mathrm{inc}}(\pi)=\sum_jr_jH_\pi(\log a_j)$. For fixed bets we compute these crossing values recursively and search numerically for the best parameter. The adaptive benchmark solves a discretized Bellman equation whose state records current evidence and the next unpaid milestone. We evaluate the adaptive and plug-in policies on $500{,}000$ simulated paths each, and  the plotted ratios have standard errors below $0.0016$. Simulation horizons make the omitted expected receipts smaller than $10^{-8}$.

Figure~\ref{fig:efficiency} uses the same Bernoulli family and a Gaussian family. For $X\sim N(\theta,\varsigma^2)$ with $\theta>0$, use $f_\lambda(X)=\exp(\lambda X-\lambda^2/8)$, $\lambda\geq0$. Then
\[
\mu(\lambda)=\lambda\theta-\lambda^2/8,
\qquad
\psi_\lambda(s)=s\mu(\lambda)+s^2\lambda^2\varsigma^2/2.
\]
Maximizing the drift gives $\lambdagro=4\theta$. Maximizing the CGF over $\lambda$ and solving the moment-root equation gives
\[
\rho^*=\frac{\log(1/\delta)}{2(\theta^2+2\varsigma^2\log(1/\delta))},
\qquad
\lambda_\cgf=4\theta+\frac{8\varsigma^2\log(1/\delta)}{\theta}.
\]
The experiments use $\varsigma=0.5$.

The Gaussian plug-in starts at $\lambda_1=1$, then uses $\widehat\lambda_t=\mathrm{clip}(4\bar X_t,0,20)$ and estimated drift $\widehat\lambda_t\bar X_t-\widehat\lambda_t^2/8$, with $\lambda_{t+1}=\widehat\lambda_t$. For the Bernoulli plug-in's deferred threshold, the estimated drift is $\mathrm{KL}(\widetilde p_t\|m)$ when $\widetilde p_t>m$, and zero otherwise. This uses the clipped fitted law, a variant of the empirical drift estimate in Section~\ref{sec:drift-policies}.

The Hoeffding baseline uses $f_\lambda(X)=\exp\{\lambda(X-m)-\lambda^2/8\}$ with $\lambda=4(p-m)$ and the drift threshold corresponding to $\mu=2(p-m)^2$. By Pinsker's inequality, its drift $2(p-m)^2$ is at most the betting drift $\mathrm{KL}(p\|m)$. For patient agents this gap carries over to utility: with $C=1$ and $\rho_\lambda\approx\log(1/\delta)/\mu(\lambda)$, the maximized value~\eqref{eq:cgf-maximized-value} is approximately $\mu(\lambda)/(e\log(1/\delta))$, which is linear in the drift. The Hoeffding baseline should therefore earn about the fraction $2(p-m)^2/\mathrm{KL}(p\|m)$ of the value of the drift rule with the betting reward. At $\delta=0.995$, this fraction falls from $0.998$ at $p=0.55$ to $0.934$ at $p=0.8$, and the simulated ratios agree with it to within $0.003$.

We obtain the deferred Bellman benchmark by policy iteration on a log-evidence grid, averaging over Gaussian increments by quadrature and summing over the two Bernoulli outcomes exactly. Each policy and the benchmark are evaluated on $20{,}000$ simulated paths under the original observation law. All paths run for at least one trial and are capped at $N=2500$, with payoff $\delta^{\tau\wedge N}\log R_{\tau\wedge N}$ even if the threshold was not reached. The ratios therefore have sampling uncertainty and Bellman discretization error.

For the large-overshoot example in Section~\ref{sec:deferred-experiments}, we compare $\lambda_\cgf\approx54.4951$ with log-evidence threshold $1/\rho^*\approx1.171936$ against $\lambda=15.5$ with exact threshold $\beta^\star(15.5)\approx0.301722$. The error relative to the uncapped values is below $2\cdot10^{-25}$.

\end{document}